\documentclass[a4paper,fleqn,usenatbib]{mnras}

\usepackage{newtxtext,newtxmath}

\usepackage[T1]{fontenc}
\usepackage{ae,aecompl}

\usepackage{graphicx}	
\usepackage{amsmath}	
\usepackage{pdflscape}  

\defcitealias{Tang2019}{T19}

\title[]{Rest-frame UV spectroscopy of extreme [OIII] emitters at $\mathbf{1.3<z<3.7}$: Toward a high-redshift UV reference sample for {\it JWST}}

\author[M. Tang et al.]{
Mengtao Tang$^{1}$\thanks{tangmtasua@email.arizona.edu}, 
Daniel P. Stark$^{1}$, 
Jacopo Chevallard$^{2}$, 
St\'{e}phane Charlot$^{2}$, \newauthor
Ryan Endsley$^{1}$ and 
Enrico Congiu$^{3,4}$
\\
\\
$^{1}$ Steward Observatory, University of Arizona, 933 N Cherry Ave, Tucson, AZ 85721, USA \\
$^{2}$ Sorbonne Universit\'{e}, UPMC-CNRS, UMR7095, Institut d'Astrophysique de Paris, F-75014, Paris, France \\
$^{3}$ Departamento de Astronom\'{i}a, Universidad de Chile, Camino del Observatorio 1515, Las Condes, Santiago, Chile \\
$^{4}$ Las Campanas Observatory - Carnegie Institution for Science, Colina el Pino, Casilla 601, La Serena, Chile \\
}

\pubyear{2020}

\begin{document}
\label{firstpage}
\pagerange{\pageref{firstpage}--\pageref{lastpage}}
\maketitle

\begin{abstract}

Deep spectroscopy of galaxies in the reionization era has revealed intense 
C~{\small III}] and C~{\small IV} line emission (EW $>15-20$~\AA). 
In order to interpret the nebular emission emerging at $z>6$, we have begun targeting 
rest-frame UV emission lines in galaxies with large specific star formation rates (sSFRs) at $1.3<z<3.7$. 
We find that C~{\small III}] reaches the EWs seen at $z>6$ only in 
large sSFR galaxies with [O~{\small III}]+H$\beta$ EW $>1500$~\AA. 
In contrast to previous studies, we find that many galaxies with intense [O~{\small III}] 
have weak C~{\small III}] emission (EW $=5-8$~\AA), suggesting that the radiation field 
associated with young stellar populations is not sufficient to power strong C~{\small III}]. 
Photoionization models demonstrate that the spread in C~{\small III}] among systems with 
large sSFRs ([O~{\small III}]+H$\beta$ EW $>1500$~\AA) is driven by variations in metallicity, 
a result of the extreme sensitivity of C~{\small III}] to electron temperature. 
We find that the strong C~{\small III}] emission seen at $z>6$ (EW $>15$~\AA) requires 
metal poor gas ($\simeq0.1\ Z_\odot$) whereas the weaker C~{\small III}] emission in our sample 
tends to be found  at moderate metallicities ($\simeq0.3\ Z_\odot$). The luminosity distribution of 
the C~{\small III}] emitters in our $z\simeq1-3$ sample presents a consistent picture, 
with stronger emission generally linked to low luminosity systems ($M_{\rm{UV}}>-19.5$) 
where low metallicities are more likely. We quantify the fraction of 
strong C~{\small III}] and C~{\small IV} emitters at $z\simeq1-3$, 
providing a baseline for comparison against $z>6$ samples. 
We suggest that the first UV line detections at $z>6$ can be explained 
if a significant fraction of the early galaxy population is found at 
large sSFR ($>200$~Gyr$^{-1}$) and low metallicity ($<0.1\ Z_\odot$).

\end{abstract}

\begin{keywords}
cosmology: observations - galaxies: evolution - galaxies: formation - galaxies: high-redshift
\end{keywords}




\section{Introduction} \label{sec:introduction}

Our understanding of galaxies in the reionization era has advanced 
considerably in the past decade following a series of multi-wavelength 
imaging surveys conducted with the {\it Hubble Space Telescope} ({\it HST}) 
\citep[e.g.][]{McLure2013,Bouwens2015,Finkelstein2015,Livermore2017,Atek2018,Oesch2018}. 
These campaigns have revealed thousands of galaxies thought to lie at $z>6$, 
providing our first census of the star-forming sources thought responsible for reionization 
(see \citealt{Stark2016,Dayal2018} for reviews). The {\it James Webb Space Telescope} ({\it JWST}) 
will soon build on this progress, delivering the first detailed spectroscopic investigation 
of $z>6$ galaxies and opening the door for investigations of stellar populations 
and the build-up of metals in early star-forming systems.

The first glimpse of the nebular emission line properties of the $z>6$ 
population has emerged in the last several years, providing a preview 
of the science that will be possible with {\it JWST}. 
Deep rest-frame ultraviolet (UV) spectra have revealed intense 
line emission from highly-ionized species of carbon 
(C~{\small III}], C~{\small IV}; e.g., 
\citealt{Stark2015a,Stark2015b,Stark2017,Laporte2017,Mainali2017,Hutchison2019}), 
indicating a hard ionizing spectrum 
that is not seen in typical star-forming galaxies at lower redshifts. 
The origin of the hard photons remains a matter of debate, with 
some suggesting AGN are required \citep{Nakajima2018} and others suggesting that 
metal poor massive star populations are sufficient \citep{Stark2017}. 
While direct constraints on the rest-frame optical spectra await {\it JWST}, 
progress has been made possible by {\it Spitzer}/IRAC broadband photometry, 
with the $[3.6]-[4.5]$ colors of $z>6$ galaxies commonly revealing 
the presence of strong [O~{\small III}]+H$\beta$ nebular emission. 
The typical rest-frame equivalent widths (EWs) at $z\simeq6-8$ 
(EW$_{\rm{[OIII]+H\beta}}\simeq600-800$~\AA; e.g., \citealt{Labbe2013,Smit2015,deBarros2019,Endsley2020}) 
indicate both strong line emission and weak underlying continuum, 
a signpost of galaxies dominated by very young stellar populations 
($<50$~Myr), as expected for systems undergoing rapidly rising star formation histories. 

Work is now underway to determine what these emission line 
properties are telling us about the reionization-era 
population. Central to these efforts are campaigns 
focused on galaxies with similar emission line properties 
at lower redshifts where they can be studied in greater detail.  
Sizeable samples of $z\simeq1-2$ galaxies with large
[O~{\small III}]+H$\beta$ EWs have been identified in broadband 
imaging and grism spectroscopy \citep[e.g.][]{Atek2011,vanderWel2011,Maseda2013,Maseda2014}, 
revealing a young population of low mass galaxies with high
densities of massive stars. In \citet[][hereafter \citetalias{Tang2019}]{Tang2019}, we built 
on these studies, targeting the strong rest-frame optical lines 
in $227$ galaxies with [O~{\small III}]+H$\beta$ EW $=300-3000$~\AA. 
The spectra revealed that the ionizing efficiency of galaxies 
(defined as the ratio of the hydrogen ionizing photon production
rate and the far-UV continuum luminosity at $1500$~\AA) increases
with the [O~{\small III}] EW. The ionization state of the
nebular gas (as probed by the [O~{\small III}]/[O~{\small II}] flux ratio, hereafter O32) 
was also found to increase with the [O~{\small III}] EW, suggesting very 
large ionization parameters in the most extreme line emitters. 
Investigations of nearby galaxies show nearly-identical trends 
\citep{Chevallard2018}, suggesting little redshift 
evolution in the trends with [O~{\small III}] EW. Taken together, these 
studies indicate that very hard spectra and highly-ionized gas 
conditions are common in a short window following a burst of star
formation in low mass galaxies. 

Progress has also been made in our understanding of the rest-frame UV 
nebular line detections. A series of {\it HST}/UV spectroscopy 
programs have investigated the stellar populations and gas conditions 
required to power C~{\small III}] and C~{\small IV} in nearby star-forming galaxies 
\citep{Senchyna2017,Senchyna2019b,Berg2019}. The results have revealed that C~{\small III}] 
is intense in galaxies with moderately 
low metallicities ($<0.4\ Z_\odot$) and large specific star formation 
rates (sSFRs), with the C~{\small III}] EW increasing in lockstep with the [O~{\small III}] EW 
\citep[e.g.][]{Rigby2015,Senchyna2017,Senchyna2019b}. Both nebular 
C~{\small IV} and He~{\small II} appear in lower metallicity galaxies ($<0.1\ Z_\odot$), 
provided the sSFR (or H$\beta$ EW) is large enough to guarantee a 
significant population of high-mass stars \citep{Senchyna2019b}. 
The local C~{\small III}] and C~{\small IV} trends have been successfully 
explained in the context of photoionization models \citep{Jaskot2016,Gutkin2016,Byler2018,Plat2019}.  
But in spite of these successes, the local studies have struggled 
to identify galaxies with C~{\small III}] and C~{\small IV} equivalent widths as 
large as seen at $z>6$, making it challenging to interpret  
the emerging body of reionization-era spectra.

Rest-frame UV surveys at $z\simeq1-4$ have provided a viable way 
forward, leading to the discovery of a significant number of 
intense UV line emitters in the last decade 
\citep{Erb2010,Stark2014,deBarros2016,Amorin2017,Maseda2017,Vanzella2017,Berg2018,LeFevre2019,Feltre2020}. 
To date, nearly all $z\simeq1-4$ 
galaxies with UV line properties similar to those seen at $z>6$ 
(i.e., EW$_{\rm{CIII]}}>20$~\AA\ or EW$_{\rm{CIV}}>20$~\AA)
appear to be AGN \citep{Nakajima2018,LeFevre2019}, 
leading to the suggestion that stellar radiation fields may be 
incapable of powering the line emission seen in the reionization 
era. Motivated by the relationship between C~{\small III}] and [O~{\small III}] EW 
\citep{Maseda2017,Du2020}, efforts  
targeting star-forming galaxies have begun to focus on very young ($<10$~Myr) 
stellar populations with the most extreme optical line emission (EW$_{\rm{[OIII]+H\beta}}>1500$~\AA), with one 
recently-discovered system identified in this manner 
displaying UV line properties that are nearly identical to those discovered in the reionization era (i.e., EW$_{\rm{CIII]}}>20$~\AA; 
\citealt{Mainali2020}). This finding suggests that stellar populations may be sufficient to power the nebular emission seen 
at $z>6$, but without larger samples, it is impossible to 
generalize to the broader population of young galaxies with 
EW$_{\rm{[OIII]+H\beta}}>1500$~\AA.

Motivated by this shortcoming, work is now underway to build 
large rest-frame UV spectral databases of galaxies with extreme optical 
line emission. The first statistical 
measures of the rest-frame UV spectral properties of galaxies selected 
to have large EW [O~{\small III}]+H$\beta$ emission have emerged 
recently \citep{Maseda2017,Du2020}. 
Here we describe results of an ongoing spectroscopic campaign 
aimed at building on this progress. In particular, we discuss
rest-frame UV spectra of $138$ extreme [O~{\small III}] emitting galaxies at $1.3<z<3.7$ 
together with a smaller sample of $4<z<6$ galaxies. Our survey includes 
$26$ galaxies with the most extreme [O~{\small III}]+H$\beta$ EWs ($>1500$~\AA), 
nearly $9\times$ greater than that in previous studies.  Our goals are twofold.
First we aim to build a large enough sample 
of $z\simeq1-3$ galaxies with [O~{\small III}]+H$\beta$ EW $=1000-2000$~\AA\ 
such that we can assess whether the C~{\small III}] intensities seen at $z>6$ 
(i.e., C~{\small III}] EW $>20$~\AA) commonly arise in galaxies with extremely 
young stellar populations, as suggested by the single 
galaxy identified in \citet{Mainali2020}. And secondly, we 
seek to quantify the rate at which very metal poor 
stellar populations ($<0.1\ Z_\odot$) appear among low 
mass extreme line emitting galaxies at $z\simeq1-3$ via  
measurement of the fraction of nebular C~{\small IV} emitters (with 
line ratios indicating a stellar origin) in our spectroscopic 
sample. This will provide a baseline for comparison against 
future measurements at $z>6$, giving the control needed to 
track the rise of low metallicity massive-star populations 
in early galaxies matched to our sample in $M_{\rm{UV}}$ and sSFR.

The organization of this paper is as follows. We describe the 
spectroscopic sample and rest-frame UV spectroscopic observations, 
as well as our photoionization modeling procedures in Section 
\ref{sec:obs_model}. We then present the rest-frame UV spectra 
of our $1.3<z<6$ galaxies in Sections \ref{sec:result}, 
and the physical properties inferred from spectra in Section \ref{sec:phy}. 
We discuss the implications of these results in Section \ref{sec:discussion} 
and summarize our conclusions in Section \ref{sec:summary}. 
We adopt a $\Lambda$-dominated, flat universe with $\Omega_{\Lambda}=0.7$,
$\Omega_{\rm{M}}=0.3$, and H$_0=70$ km s$^{-1}$ Mpc$^{-1}$. All
magnitudes in this paper are quoted in the AB system \citep{Oke1983},
and all equivalent widths are quoted in the rest-frame. 


\begin{figure*}
\begin{center}
\includegraphics[width=0.95\linewidth]{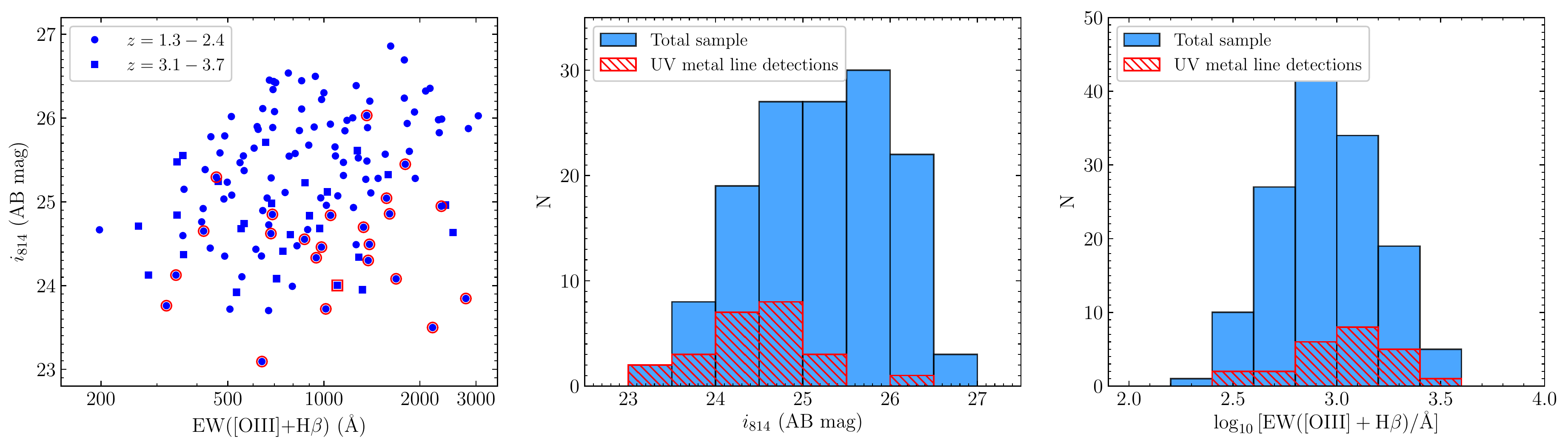}
\caption{{\it HST} F814W magnitude and [O~{\scriptsize III}]+H$\beta$ EW distribution of the $138$ extreme [O~{\scriptsize III}] emitters at $z=1.3-3.7$ in our spectroscopic sample. Left panel: F814W magnitudes and [O~{\scriptsize III}]+H$\beta$ EWs of the $138$ galaxies (objects at $z=1.3-2.4$ are shown by blue circles, and objects at $3.1-3.7$ are shown by blue squares). Those sources with UV metal line detections are  additionally demarcated by larger red circles or squares. Middle panel: F814W magnitude distributions of the total extreme [O~{\scriptsize III}] emitter sample (solid blue) and the subset with UV metal emission line detections presented in Section \ref{sec:spec_z14} (dashed red). Right panel: [O~{\scriptsize III}]+H$\beta$ EW distributions of the total sample (solid blue) and the subset with UV metal line detections (dashed red).}
\label{fig:mag}
\end{center}
\end{figure*}


\section{Observations and analysis} \label{sec:obs_model}

In this work, we aim to investigate the rest-frame UV spectra of galaxies with extreme EW optical line emission (extreme emission line galaxies, EELGs). We primarily focus on the extreme [O~{\small III}] line emitting galaxies at $z=1.3-3.7$, obtaining rest-frame UV spectra via ground-based optical spectrographs. In addition, we present near-infrared spectroscopy of a subset of sources at $z\sim4-6$. The spectroscopic observations of each sample are described in Section \ref{sec:obs_z14} and Section \ref{sec:obs_z46}, respectively. We then describe the photoionization modeling we apply to our sample in Section \ref{sec:modeling}.

\subsection{Optical spectroscopy of $z=1.3-3.7$ EELGs} \label{sec:obs_z14}

The rest-frame UV spectra presented in this subsection build on a large spectroscopic effort to obtain rest-frame optical spectra of extreme [O~{\small III}] emitters at $z=1.3-3.7$ in the CANDELS fields (\citetalias{Tang2019}; Tang et al. in prep) and the recent rest-frame UV spectroscopic surveys presented in \citet{Du2020} and \citet{Mainali2020}. We direct the reader to \citetalias{Tang2019} for the full description of the sample selection and the follow-up near-infrared (rest-frame optical) spectroscopic observations. The extreme [O~{\small III}] emitters are required to have large rest-frame [O~{\small III}]~$\lambda\lambda4959,5007$ EWs with values $\simeq300-2000$~\AA, which are chosen to match the range expected to be common in reionization-era galaxies \citep[e.g.][]{Endsley2020}. In order to measure the rest-frame UV emission lines of extreme [O~{\small III}] emitters at $z=1.3-3.7$, we have been conducting optical spectroscopic observations. Data were taken over three observing runs between 2018 and 2019 with optical spectrographs on the Magellan telescopes and the MMT.

The majority of our optical spectroscopic follow-up program is conducted using the
Inamori-Magellan Areal Camera \& Spectrograph (IMACS; \citealt{Dressler2011}) on 
the Magellan Baade telescope. We use IMACS in multi-slit spectroscopy mode. We 
designed two masks in the Cosmic Evolution Survey (COSMOS) field and 
the Ultra Deep Survey (UDS) field utilizing the `maskgen'
software, targeting $114$ extreme [O~{\small III}] emitters
(EW$_{\rm{[OIII]}\lambda\lambda4959,5007}>300$~\AA) at $z\simeq1.3-3.7$. The targets
were placed on the IMACS masks using a similar selection function introduced in
Section 2.2 of \citetalias{Tang2019}. We adjust the target priority based on the
[O~{\small III}] EWs inferred from {\it HST} grism spectra (at $z=1.3-2.4$;
\citetalias{Tang2019}) or the $K$-band flux excess (at $z=3.1-3.7$; Tang et al. in
prep). Objects with the largest EWs ([O~{\small III}] EW $>1000$~\AA) are very
rare, and thus we give the highest priority to the largest EW [O~{\small III}]
line emitting galaxies. In addition, we are particularly interested in comparing
the properties derived from rest-frame UV spectroscopy and those from
rest-frame optical spectroscopy. Therefore, we increase the priority of objects
whose rest-frame optical spectra have been taken from our near-infrared
spectroscopic follow-up program (\citetalias{Tang2019}; Tang et al. in prep). 

The details of the IMACS observations are summarized in Table \ref{tab:opt_obs}. 
We used the $300$ lines/mm grism blazed at $17.5$ degrees on the f/2 camera. The
grism provides wavelength coverage from $3900$~\AA\ to $10000$~\AA, covering UV
metal lines including C~{\small IV}, O~{\small III}], and 
C~{\small III}], as well as Ly$\alpha$ emission line at $z>2.2$ and 
[O~{\small II}]~$\lambda\lambda3727,3729$ emission lines at $z<1.68$. The 
slit length and width of our IMACS mask were set to $6.0$~arcsec and $1.0$~arcsec,
respectively. The $1.0$~arcsec slit width results in a spectral resolution of 
$6.7$~\AA\ for the entire wavelength coverage. For the mask targeting the COSMOS
field, we placed $60$ extreme [O~{\small III}] emitters at $z\simeq1.3-3.7$
with [O~{\small III}]~$\lambda\lambda4959,5007$ EW $\simeq300-2000$~\AA. The
targets have $i_{814}$-band magnitude from $23.1$ to $26.7$, with a median value of
$25.2$. We observed this mask on 2019 March 06 and 07 for a total on-source
integration time of 5.7 hours with an average seeing of $0.8$~arcsec. We placed $54$
extreme [O~{\small III}] emitters at $z\simeq1.3-3.7$ with 
[O~{\small III}]~$\lambda\lambda4959,5007$ EW $\simeq300-1600$~\AA\ on the 
mask targeting the UDS field. The $i_{814}$-band magnitudes of these targets have a
range from $23.8$ to $26.9$ (median $i_{814}=25.5$). This mask was observed on 2019
October 30 for an integration time of $3$ hours with seeing of $0.8$~arcsec. Each
mask contains two slits stars to compute the absolute flux calibration. We also
observed spectrophotometric standard stars at a similar airmass in each observing
run in order to correct the instrumental response.


\begin{figure*}
\begin{center}
\includegraphics[width=0.95\linewidth]{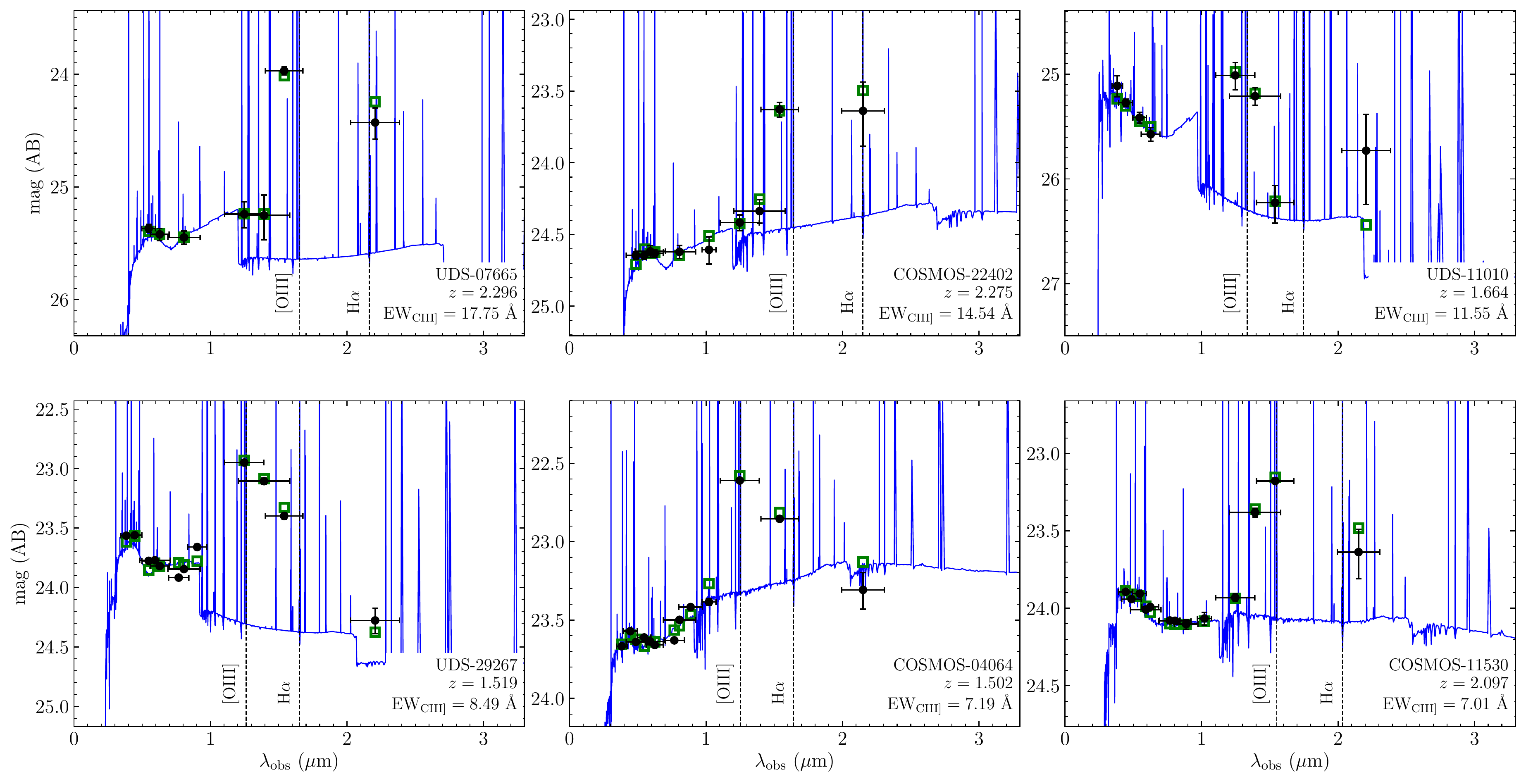}
\caption{Examples of the broadband SEDs of $z=1.3-3.7$ extreme 
[O~{\scriptsize III}] emitters with rest-frame UV metal line detections. 
Observed broadband photometry is shown as solid black circles. The best-fit SED 
models inferred from BEAGLE are plotted by solid blue lines, and synthetic
photometry is presented by open green squares.}
\label{fig:sed}
\end{center}
\end{figure*}

We reduced the IMACS spectra using the publicly available data reduction pipeline 
Carnegie Observatories System for MultiObject Spectroscopy\footnote{\url{https://code.obs.carnegiescience.edu/cosmos}\\
The acronyms of the IMACS data reduction pipeline Carnegie Observatories System for MultiObject Spectroscopy, COSMOS, is incidentally the same as that of the Cosmic Evolution Survey. When mentioning COSMOS throughout the paper, we only refer to the Cosmic Evolution Survey.} 
\citep{Dressler2011,Oemler2017}. During the observing runs, flat fields were 
obtained using quartz high lamps, and arcs were obtained using HeNeAr lamps. The pipeline 
performs bias subtraction, flat-fielding, wavelength calibration, sky subtraction, and
extracts the two-dimensional (2D) spectra. We created one-dimensional (1D) spectra from the
reduced 2D spectra using a boxcar extraction, with the extraction aperture matched to the
spatial profile of the object. We derived the transmission curve from the observed standard
star spectra and corrected the instrumental response. Slit loss corrections were performed
following the similar procedures in \citetalias{Tang2019} (see also \citealt{Kriek2015}).
We derived the spatial profile of each target from its {\it HST} F606W postage stamp, and
computed the fraction of the light within the slit to that of the total spatial profile.
The flux of each spectrum was then divided by the in-slit light fraction measured for each
object. Finally, the absolute flux calibration was performed using the slit stars, by 
comparing the slit-loss corrected count rates of slit star spectra with the flux in the
\citet{Skelton2014} catalogs. For objects on the mask targeting the COSMOS field, the
median $3\sigma$ line flux sensitivity of the IMACS spectra is $3.6\times10^{-18}$~erg~s$^{-1}$~cm$^{-2}$ 
(assuming a line width of $13.4$~\AA, i.e., $2\times$ the spectral 
resolution) in the wavelength range ($4800-6300$~\AA) where C~{\small III}] 
(at $z\sim2$) or Ly$\alpha$ (at $z\sim3$) is situated. This flux limit enables us to detect
emission lines ($3\sigma$) with rest-frame EW $\sim1$, $4$, and $17$~\AA\ for the sources
with the brightest ($23.1$), median ($25.2$), and the faintest ($26.7$) continuum
magnitudes. For objects on the mask targeting the UDS field, we reached a median
$3\sigma$ line flux sensitivity of $6.9\times10^{-18}$~erg~s$^{-1}$~cm$^{-2}$ in
$4800-6300$~\AA. This sensitivity provides $3\sigma$ line EW limits of $\sim2$, $11$, and
$41$~\AA\ for the sources with the brightest ($23.8$), median ($25.5$), and the faintest
($26.9$) continuum magnitudes.


\begin{table*}
\begin{tabular}{|c|c|c|c|c|c|c|c|}
\hline
Field & Number of Targets & R.A. & Decl. & P.A. & Instrument & Exposure Time & Average Seeing \\
 & & (hh:mm:ss) & (dd:mm:ss) & (deg) & & (hours) & (arcsec) \\
(1) & (2) & (3) & (4) & (5) & (6) & (7) & (8) \\
\hline 
\hline
COSMOS & $60$ & 10:00:21 & 02:18:00 & $-$80 & Magellan/IMACS & 5.7 & 0.8 \\
UDS & $54$ & 02:17:26 & $-$05:12:13 & $-$8 & Magellan/IMACS & 3.0 & 0.8 \\
UDS & $24$ & 02:17:26 & $-$05:05:40 & $-$86 & MMT/Binospec & 5.0 & 1.2 \\
\hline
\end{tabular}
\caption{Summary of optical spectroscopic observations of extreme [O~{\scriptsize III}]
emitters at $z=1.3-3.7$. Column (1): CANDELS field; Column (2): Number of science targets
on each mask, alignment stars and slit stars are not included; Column (3): Right ascension
of the mask center; Column (4): Declination of the mask center; Column (5): Position angle
of the mask; Column (6): Instrument being used; Column (7): Total exposure time of each
observing run; Column (8): Average seeing during the observation, in full width at half
maximum.}
\label{tab:opt_obs}
\end{table*}

We also obtained optical spectra from the Binospec \citep{Fabricant2019} on the MMT with
multi-slit spectroscopy mode. We designed one mask in the UDS field using the `BinoMask'
software. Twenty-four extreme [O~{\small III}] emitters ([O~{\small III}]~$\lambda\lambda4959,5007$ 
EW $\simeq700-2500$~\AA) were placed on the mask with the same
selection function used in Magellan/IMACS observations. The targets have $i_{814}$
magnitude from $23.7$ to $26.7$ (median $i_{814}=25.7$). Spectra were taken using the $270$
lines/mm grism blazed at $5.5$ degrees, with wavelength coverage from $3850$~\AA\ to $9000$~\AA. 
The slit width was set to $1.0$~arcsec, which results in a spectral resolution of
$R=1340$. We observed this mask on 2018 October 18 and 2018 November 03 for a total
integration time of $5$ hours with an average seeing of $1.2$~arcsec. The mask contains 2
slit stars to derive absolute flux calibration, and standard stars were observed to correct
the instrumental response. We summarize the details of the Binospec observations in Table
\ref{tab:opt_obs}.

We reduced the Binospec spectra using the publicly available data reduction pipeline\footnote{\url{https://bitbucket.org/chil\_sai/binospec}} \citep{Kansky2019}. The pipeline performs flat-fielding, wavelength calibration, and sky subtraction. Before extracting the 2D spectra, the pipeline corrects atmospheric extinction and instrumental response using the sensitivity curve derived from observations of standard stars. We extracted the 1D spectra using a boxcar extraction. The absolute flux calibration was performed using slit stars, and slit loss correction was performed following the same procedures used in Magellan/IMACS observations. The median $3\sigma$ line flux sensitivity of our Binospec spectra is $7.2\times10^{-18}$~erg~s$^{-1}$~cm$^{-2}$ in $4800-6300$~\AA, providing $3\sigma$ line EW limits of $\sim2$, $14$, and $34$~\AA\ for the sources with the brightest ($23.7$), median ($25.7$), and the faintest ($26.7$) continuum magnitudes. 

Our observations described above obtained optical (rest-frame UV) spectra for $138$ 
extreme [O~{\small III}] emitters at $z=1.3-3.7$. The $i_{814}$ magnitude 
distribution of these $138$ galaxies are shown in Figure \ref{fig:mag}. We also add
available near-infrared (rest-frame optical) spectra which provide useful constraints on
the ionizing radiation fields and the interstellar medium (ISM) conditions. We have obtained ground-based
near-infrared spectra for $33$ of the $138$ sources (\citetalias{Tang2019}; Tang et al. in
prep), with strong [O~{\small III}]~$\lambda5007$ or H$\alpha$ emission lines detected
at S/N $>5$. Among these thirty-three objects, twenty-four have the full suite of strong rest-frame
optical emission line detections ([O~{\small II}], H$\beta$, [O~{\small III}],
H$\alpha$), allowing us to measure ionization and metallicity sensitive line ratios 
(i.e., [O~{\small III}]/[O~{\small II}]). For the remaining objects without
ground-based near-infrared spectroscopic observations, we utilize the H$\beta$, 
[O~{\small III}], or H$\alpha$ emission lines measured from {\it HST} grism
near-infrared spectra (\citealt{Momcheva2016}) if the grism S/N is larger than $5$. This
results in a sample of $116$ (of $138$) galaxies with [O~{\small III}] flux
measurements. We then compute [O~{\small III}] EWs following \citetalias{Tang2019} 
and (Tang et al. in prep). In Section \ref{sec:discussion}, we will discuss the 
correlations between rest-frame UV spectral properties and rest-frame optical properties 
(i.e., [O~{\small III}]~$\lambda5007$ EW). 

\subsection{Near-infrared spectroscopy of $z\sim4-6$ EELGs} \label{sec:obs_z46}

We have additionally obtained rest-frame UV spectra for a smaller number of extreme
optical line emitter candidates at $z\sim4-6$. For galaxies in this redshift range,
strong rest-frame optical emission lines (H$\alpha$, [O~{\small III}]) are
situated in {\it Spitzer}/IRAC $3.6\ \mu$m or $4.5\ \mu$m band, allowing us to
identify extreme optical line emitters via large flux excesses in the $[3.6]$ or
$[4.5]$ filters.  Recent studies \citep[e.g.][]{Shim2011,Stark2013,Rasappu2016} reveal
that extreme H$\alpha$ emitters at $3.8<z<5.0$ ($5.1<z<5.4$) show blue (red)
$[3.6]-[4.5]$ colors since the H$\alpha$ emission line is within the $[3.6]$ ($[4.5]$)
filter while the $[4.5]$ ($[3.6]$) filter is dominated by rest-frame optical
continuum. For galaxies at $5.4<z<6.0$, [O~{\small III}]+H$\beta$ and
H$\alpha$+[N~{\small II}] emission lines are situated in $[3.6]$ and $[4.5]$
filters, respectively. In this subsection, we first describe selection of a 
sample of extreme optical line emitter candidates at $z\sim4-6$ using  
IRAC color excess methods, then describe the near-infrared (rest-frame UV) 
spectroscopic observations of a subset of the sources in this sample.

Our $z\sim4-6$ targets were selected from public samples of Lyman break galaxies (LBGs) 
with spectroscopic or photometric redshift measurements. Taking advantage of 
the spectroscopic redshifts provided in \citet{Stark2013} and \citet{Rasappu2016}, we
identified a subset of sources at $3.8<z_{\rm{spec}}<5.0$ and $5.1<z_{\rm{spec}}<5.4$.
Redshifts in the above papers are derived from Ly$\alpha$ emission lines or UV
absorption lines. We also selected LBGs with photometric redshifts at $3.8<z<6.0$ (and
with $1\sigma$ photo-$z$ uncertainty $\delta z<0.1$ to ensure robust redshift
estimates) from \citet{Bouwens2015}, where the redshifts are computed using EAZY
\citep{Brammer2008}. To ensure self-consistent astrometry, we matched the coordinates
of all the sources with spectroscopic or photometric redshifts to \citet{Skelton2014}
catalogs. To efficiently identify extreme emission line galaxies, we required
$[3.6]-[4.5]<-0.3$ for sources at $3.8<z<5.0$ and $[3.6]-[4.5]>0.3$ for sources at
$5.1<z<5.4$. We also required S/N $>5$ for the $[3.6]$ and $[4.5]$ fluxes so that the
H$\alpha$ EWs can be accurately derived from IRAC colors. The above $[3.6]-[4.5]$
color cuts correspond to H$\alpha$+[N~{\small II}]+[S~{\small II}] EW $>500$~\AA, 
which is greater than the average H$\alpha$ EW estimated for star-forming
galaxies in the reionization era \citep{Labbe2013}. For sources at $5.4<z<6.0$, 
we do not apply an IRAC color cut since strong rest-frame optical emission lines 
are within both $[3.6]$ and $[4.5]$ filters. 

We use the catalog of $z\sim4-6$ extreme optical line emitter candidates described above as input for our near-infrared spectroscopic observations. The spectra presented in this paper were obtained using the multi-object spectrograph MOSFIRE \citep{McLean2012} on the Keck I telescope. We observed two masks on 2015 November 30, targeting six sources in the GOODS-S field and five sources in the COSMOS field. The remainder of the mask was filled with sources discussed in previous publications (\citealt{Stark2017}; \citetalias{Tang2019}). The slit widths were set to $0.7$~arcsec, resulting in a spectral resolution of $R=3388$ ($\Delta\lambda\simeq2.9$~\AA\ at the blue end, $\sim9716$~\AA, and $\Delta\lambda\simeq3.3$~\AA\ at the red end, $\sim11250$~\AA). 

For the mask targeting the GOODS-S field, we observed six objects at $3.8<z<5.0$. The primary targets of this mask are two extreme H$\alpha$ emitting candidates, GOODS-S-46692 and GOODS-S-41253. GOODS-S-46692 is a bright galaxy ($J_{125}=24.9$) with a spectroscopic redshift derived from Ly$\alpha$ emission, $z_{\rm{Ly}\alpha}=4.811$ \citep{Vanzella2008}, and IRAC color $[3.6]-[4.5]=-0.33$. GOODS-S-41253 is fainter ($J_{125}=26.0$) with photometric redshift of $z_{\rm{phot}}=4.28$ and IRAC color $[3.6]-[4.5]=-0.45$. We also added four sources ($J_{125}=25.2-25.7$) with photometric redshifts in the range $z_{\rm{phot}}=4.1-4.8$ and less extreme IRAC colors ($[3.6]-[4.5]=-0.1$ to $-0.3$) as fillers of this mask. We observed in $Y$-band, for a total on-source integration time of $3.2$ hours with an average seeing of $0.94$~arcsec. The spectral wavelength coverage is from $\sim9716$~\AA\ to $\sim11250$~\AA, allowing constraints on C~{\small III}]~$\lambda\lambda1907,1909$ for the chosen targets. 

For the mask targeting the COSMOS field, we observed five objects with photometric redshifts 
at $3.8<z_{\rm{phot}}<6.0$. Two bright targets ($J_{125}=24.1-24.6$) have photometric 
redshifts in the range $z=4.5-4.7$ with blue IRAC colors $[3.6]-[4.5]=-0.30$ to $-0.72$. The
other three targets ($J_{125}=24.9-26.2$) have photometric redshifts ($z=5.6-5.8$) where
both IRAC filters are contaminated by emission lines.  
We observed the COSMOS mask in $Y$-band for a total on-source integration time of $2.4$ 
hours with an average seeing of $0.74$~arcsec. The observations enable constraints on the 
C~{\small III}]~$\lambda\lambda1907,1909$ line strengths for  targets at $z=4.5-4.7$ 
and O~{\small III}]~$\lambda\lambda1661,1666$ for those at $z=5.6-5.8$. Both of the 
masks were filled with $z\simeq2$ extreme [O~{\small III}] line emitting galaxies, which
have been discussed in \citetalias{Tang2019}. We also placed two slit stars on each mask for
absolute flux calibration, and we obtained the telluric star spectrum in order to correct the
atmosphere absorption and instrument response. For both masks we performed a dither pattern 
of $\pm1.5$~arcsec along the slit for sky subtraction. The near-infrared spectroscopic observations
of $z\sim4-6$ extreme optical line emitter candidates are summarized in Table \ref{tab:nir_obs}.

The spectra were reduced using the publicly available MOSFIRE Data Reduction Pipeline (DRP\footnote{\url{https://keck-datareductionpipelines.github.io/MosfireDRP}}). The DRP performs flat-fielding, wavelength calibration, and background subtraction before extracting 2D spectra. We corrected the atmosphere absorption and instrument response using the spectrum of a telluric star, and performed slit loss correction the same way as described in  \citetalias{Tang2019} (see also \citealt{Kriek2015}). The absolute flux calibration was performed using the spectra counts and photometric flux of slit stars. We extracted 1D spectra using boxcar extraction with an aperture matched to the spatial profile of the object. Assuming a line width of $6-7$~\AA\ ($2\times$ the spectral resolution), the median $5\sigma$ line flux sensitivity is $2.6$ ($3.2$) $\times10^{-18}$~erg~s$^{-1}$~cm$^{-2}$ for spectra on the mask targeting the GOODS-S (COSMOS) field. We will present the results of our near-infrared spectroscopy of $z\sim4-6$ targets in Section \ref{sec:spec_z46}.


\begin{table*}
\begin{tabular}{|c|c|c|c|c|c|c|c|c|}
\hline
Target & $z_{\rm{Ly}\alpha}$ & $z_{\rm{phot}}$ & R.A. & Decl. & $J_{125}$ & $[3.6]-[4.5]$ & Exposure Time & Average Seeing \\
 & & & (hh:mm:ss) & (dd:mm:ss) & (mag) & (mag) & (hours) & (arcsec) \\
(1) & (2) & (3) & (4) & (5) & (6) & (7) & (8) & (9) \\
\hline 
\hline
COSMOS-18502 & ... & $4.58$ & 10:00:16.82 & 02:22:03.4 & $24.6$ & $-0.30$ & $2.4$ & $0.74$ \\
COSMOS-19732 & ... & $4.70$ & 10:00:17.80 & 02:22:46.8 & $24.1$ & $-0.72$ & $2.4$ & $0.74$ \\
COSMOS-11116 & ... & $5.62$ & 10:00:24.48 & 02:17:33.1 & $25.7$ & ... & $2.4$ & $0.74$ \\
COSMOS-15365 & ... & $5.64$ & 10:00:22.88 & 02:20:10.8 & $24.9$ & ... & $2.4$ & $0.74$ \\
COSMOS-20187 & ... & $5.73$ & 10:00:16.56 & 02:23:02.6 & $26.2$ & ... & $2.4$ & $0.74$ \\
GOODS-S-36712 & ... & 4.74 & 03:32:10.32 & $-$27:44:25.4 & $25.3$ & $-0.28$ & $3.2$ & $0.94$ \\
GOODS-S-38450 & ... & 4.58 & 03:32:09.06 & $-$27:43:51.8 & $25.2$ & $-0.20$ & $3.2$ & $0.94$ \\
GOODS-S-39157 & ... & 4.17 & 03:32:09.32 & $-$27:43:38.8 & $25.4$ & $-0.08$ & $3.2$ & $0.94$ \\
GOODS-S-40887 & ... & 4.50 & 03:32:13.24 & $-$27:43:08.4 & $25.7$ & $-0.17$ & $3.2$ & $0.94$ \\
GOODS-S-41253 & ... & 4.28 & 03:32:11.30 & $-$27:43:01.3 & $26.0$ & $-0.45$ & $3.2$ & $0.94$ \\
GOODS-S-46692 & $4.811$ & 4.66 & 03:32:10.03 & $-$27:41:32.6 & $24.9$ & $-0.33$ & $3.2$ & $0.94$ \\
\hline
\end{tabular}
\caption{Summary of near-infrared spectroscopic observations of extreme optical line emitters at $z\sim4-6$. Column (1): Target ID; Column (2): Spectroscopic redshift derived from Ly$\alpha$ emission line; Column (3): Photometric redshift from \citet{Skelton2014}; Column (4): Right ascension of the mask center; Column (5): Declination of the mask center; Column (6): $J$-band ({\it HST} F125W) magnitude; Column (7): $[3.6]-[4.5]$ colors for galaxies at $3.8<z<5.0$, which are used to estimate H$\alpha$ EW; Column (9): On-source integration time; Column (9): average seeing during the observation, in full width at half maximum.}
\label{tab:nir_obs}
\end{table*}

\subsection{Photoionization modeling} \label{sec:modeling}

We utilize photoionization models to investigate the physical properties of the extreme emission line galaxies in our rest-frame UV spectroscopic sample. We fit the broadband photometry and observed emission lines from our objects using the Bayesian SED modeling and interpreting tool BEAGLE (version 0.23.0; \citealt{Chevallard2016}). The broadband photometry is obtained from the 3D-HST catalogs \citep{Skelton2014}, and we use the multi-wavelength data covering $0.3-2.5\ \mu$m. For each galaxy, we remove fluxes in filters that lie blueward of Ly$\alpha$ to avoid introducing uncertain contributions from Ly$\alpha$ emission and Ly$\alpha$ forest absorption. We simultaneously fit the available rest-frame UV (C~{\small IV}, O~{\small III}], C~{\small III}]; see Section \ref{sec:result}) and optical ([O~{\small II}], H$\beta$, [O~{\small III}], H$\alpha$) emission line fluxes. BEAGLE adopts the photoionization models of star-forming galaxies presented in \citet{Gutkin2016}, which combines the latest version of \citet{Bruzual2003} stellar population synthesis (SPS) models with the photoionization code {\small CLOUDY} \citep{Ferland2013} to describe the emission from stars and interstellar gas. The SPS models, which are described in more detail in \citet[][see their Section 2.1]{Vidal-Garcia2017}, incorporate updated stellar-evolution calculation by \citet{Bressan2012} for stars with initial masses up to $350$ M$_{\odot}$ \citep{Chen2015}, combined with libraries of stellar spectra from various sources. In fitting the data, we will explore whether the models powered by stars can reproduce the emission line fluxes and SEDs of the sources in our spectroscopic sample.  The fits will constrain the stellar population parameters (e.g., stellar mass, age, and sSFR) as well as the ionized gas properties (e.g., metallicity, ionization parameter). 

We assume a constant star formation history, parameterizing the maximum stellar age in the range from $1$~Myr to the age of the Universe at the given redshift. For galaxies with spectroscopic redshifts measured from rest-frame UV or optical spectra, we fix the redshift of each object to its spectroscopic redshift. For sources with photometric redshift measurements only, we allow the model galaxy redshift to vary in the $1\sigma$ confidence interval of photo-$z$ for each object. We adopt a \citet{Chabrier2003} initial mass function and allow the metallicity to vary in the range $-2.2\le\log{(Z/Z_\odot)}\le0.25$ ($Z_\odot=0.01524$; \citealt{Caffau2011}). The interstellar metallicity is assumed to be the same as the stellar metallicity ($Z_{\rm{ISM}}=Z_{\star}$) for each galaxy. We consider models with an electron density of $n_e=100$ cm$^{-3}$, which is consistent with the average density inferred from typical star-forming galaxies at $z\sim2$ \citep[e.g.][]{Sanders2016,Steidel2016}. We adjust the ionization parameter $U$ and the dust-to-metal mass ratio $\xi_{\rm{d}}$ in the range $-4.0\le\log{U}\le-1.0$ and $0.1\le\xi_{\rm{d}}\le0.5$, respectively. We also consider models with a set of carbon-to-oxygen abundance ratios (C/O $=0.10$, $0.27$, $0.52$, $0.72$, and $1.00$ (C/O)$_{\odot}$, where the solar value (C/O)$_{\odot}=0.44$). To account for the effect of dust attenuation in the neutral ISM, we assume the \citet{Calzetti2000} extinction curve. We adopt an exponential distribution prior on the $V$-band dust attenuation optical depths, fixing the fraction of attenuation optical depth arising from dust in the diffuse ISM to $\mu=0.4$ \citep{Chevallard2016}. Finally, we adopt the prescription of \citet{Inoue2014} to include the absorption of IGM.

\begin{figure*}
\begin{center}
\includegraphics[width=\linewidth]{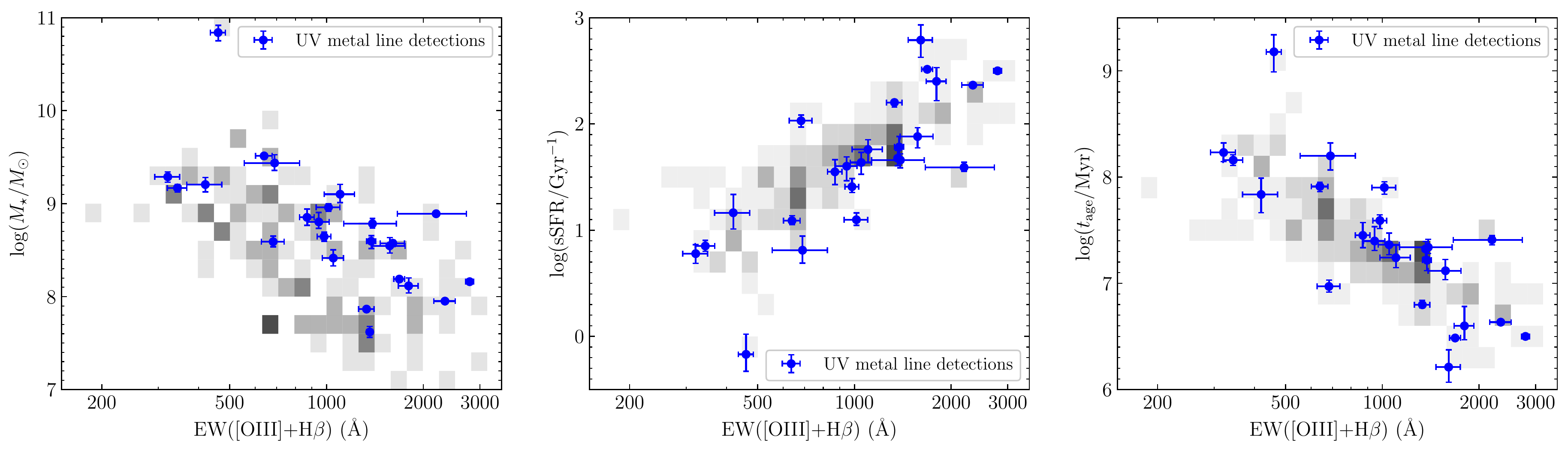}
\caption{Stellar mass (left panel), specific star formation rate (middle), and stellar age (right) as a function of [O~{\scriptsize III}]+H$\beta$ EW for the $138$ extreme [O~{\scriptsize III}] emitters at $z=1.3-3.7$ in our spectroscopic sample. The properties are derived from photoionization modeling using BEAGLE. The shaded regions represents the number of galaxies in bins of [O~{\scriptsize III}]+H$\beta$ EW and stellar mass, sSFR, or age, with a darker color indicates a larger number. We overplot the $24$ galaxies with rest-frame UV line detections in our sample (see Section \ref{sec:spec_z14}) with blue circles.}
\label{fig:ew_star}
\end{center}
\end{figure*}

With the above parameterization, we fit the SEDs and available emission line
constraints for the $z\sim1-6$ EELGs in our sample using the BEAGLE tool. 
For each free parameter described above, we adopt the median of the posterior
probability distribution as the best-fit value. 
 In Figure \ref{fig:sed}, we overplot the best-fit BEAGLE models on the observed broadband SEDs. 
The models are able to reproduce the observed SEDs and nebular emission line fluxes. 
We show the best-fit stellar population properties (stellar mass, sSFR, and stellar age) of 
the $138$ extreme [O~{\small III}] emitters at $z=1.3-3.7$ in Figure \ref{fig:ew_star}, highlighting the distribution for the subset of sources with UV line detections. 
Similar to what has been demonstrated in \citetalias{Tang2019}, galaxies with 
largest [O~{\small III}] EWs have the lowest stellar masses, the largest sSFRs, 
and the youngest stellar ages. Objects with EW$_{\rm{[OIII]+H}\beta}=1500-3000$~\AA\ show 
much larger sSFRs (median value of $218$~Gyr$^{-1}$) and younger luminosity-weighted stellar ages 
(median of $5$~Myr, assuming a constant star formation history) than those with 
EW$_{\rm{[OIII]+H}\beta}=600-800$~\AA\ 
(median sSFR $=26$~Gyr$^{-1}$ and median age $=37$~Myr). 
The latter subset corresponds to the average EW of $z\sim7-8$ galaxies  \citep{Labbe2013,deBarros2019,Endsley2020}, whereas the more 
extreme optical line emitters are  present in significant 
numbers at $z>6$ \citep{Smit2014,Smit2015,Endsley2020}, often associated 
with strong rest-frame UV line emission \citep{Stark2017,Hutchison2019}.
Here we note that the young ages 
only refer to the stellar populations that are dominating the rest-frame UV and optical SED 
and do not exclude the presence of an underlying older 
stellar population.

The underlying rest-frame UV 
continuum predicted by BEAGLE is used to compute the UV emission line EWs 
for the objects without reliable continuum measurement from spectra 
as described in Section \ref{sec:result}. For galaxies at $z\sim4-6$ in our
spectroscopic sample, BEAGLE can successfully reproduce the large IRAC flux
excess caused by nebular emission. We infer the H$\alpha$ and 
[O~{\small III}]~$\lambda5007$ EWs for our $z\sim4-6$ sources from BEAGLE
models since their rest-frame optical spectra are not visible with current
facilities. We will discuss the constraints inferred from photoionization
modeling in Section \ref{sec:phy}. 


\section{Results} \label{sec:result}

Here we describe nebular emission line constraints made possible by the rest-frame UV spectra we have obtained for our extreme emission line galaxies at $z=1.3-3.7$ (Section \ref{sec:spec_z14}) and those we have observed at $z\sim4-6$ (Section \ref{sec:spec_z46}).

\subsection{Rest-frame UV spectra at $z=1.3-3.7$} \label{sec:spec_z14}


\begin{figure*}
\begin{center}
\includegraphics[width=\linewidth]{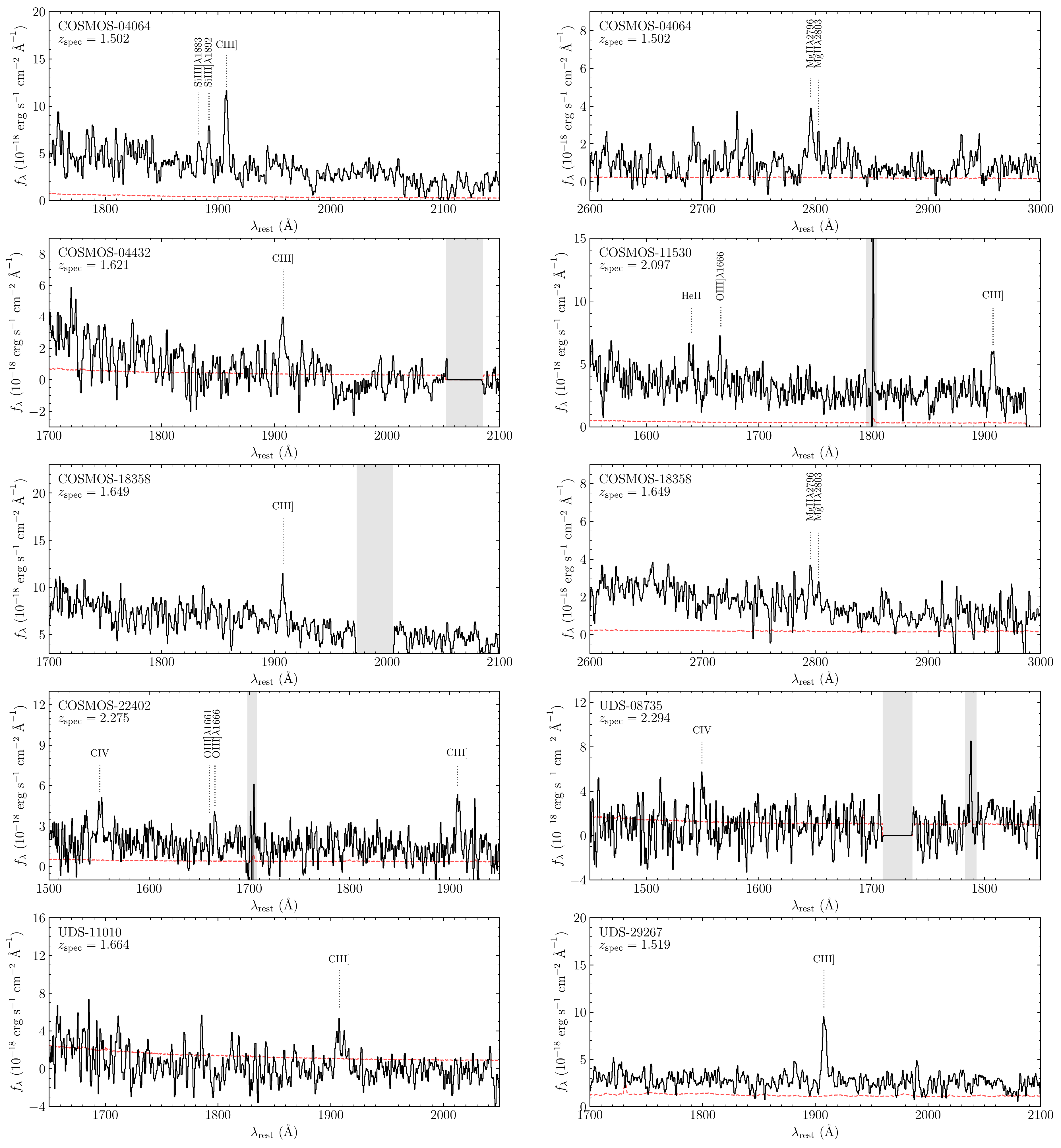}
\caption{Examples of emission lines detected in the rest-frame UV spectra of $z=1.3-3.7$ extreme [O~{\scriptsize III}] emitters. The blended C~{\scriptsize III}]~$\lambda1908$ doublet is often the most prominent emission line, and we also detected the blended C~{\scriptsize IV}~$\lambda1549$, He~{\scriptsize II}~$\lambda1640$, O~{\scriptsize III}]~$\lambda1661,1666$, Si~{\scriptsize III}]~$\lambda1883,1892$, and Mg~{\scriptsize II}~$\lambda\lambda2796,2803$ emission lines in a subset of sources. The black solid lines and red dashed lines represent flux and error, respectively. Detected emission lines are marked by black dotted lines. The grey regions indicate gaps between spectra or wavelength ranges contaminated by sky line residuals.}
\label{fig:uv_spec}
\end{center}
\end{figure*}

We characterize the rest-frame UV emission line strengths for the $138$ extreme [O~{\small III}] emitters in our $z=1.3-3.7$ sample described in Section \ref{sec:obs_z14}. The redshifts of the majority ($116$) of these $138$ sources were computed by fitting the [O~{\small III}]~$\lambda5007$ emission line from the ground-based \citepalias{Tang2019} or HST grism-based \citep{Momcheva2016} near-infrared spectra. For a subset ($22$) of objects at $z=3.1-3.7$ identified by $K_{\rm{s}}$-band excess (see Section \ref{sec:obs_z14}), we do not yet have a rest-frame optical spectroscopic redshift. For these sources, we use the Ly$\alpha$ line to compute the redshift if it is available, and otherwise we rely on the photometric redshift measurements from the \citet{Skelton2014} catalogs. Using the measured redshifts, we search for rest-frame UV emission lines including the blended C~{\small IV}~$\lambda1549$ doublet, He~{\small II}~$\lambda1640$, O~{\small III}]~$\lambda\lambda1661,1666$, Si~{\small III}]~$\lambda\lambda1883,1892$, the blended C~{\small III}]~$\lambda1908$ doublet, and Mg~{\small II}~$\lambda\lambda2796,2803$.

Emission line fluxes are determined from the extracted 1D spectra. If an emission line feature is well-detected (S/N $>5$), we apply a Gaussian fit to the line profile to derive the flux. Otherwise we calculate the line flux using direct integration. If the flux is measured with S/N $<3$, we consider the line as undetected and derive a $3\sigma$ upper limit. The line flux  limit is computed by summing the error spectrum in quadrature over $\sim200$ km/s ($14$~\AA\ for a single emission line, and $20$~\AA\ for the blended C~{\small IV} or C~{\small III}] doublet). This integration range is chosen to be consistent with the upper bound of line widths found for UV metal lines \citep[e.g.][]{Bayliss2014,James2014,Stark2014,Mainali2017}. For each detected line, we calculate the corresponding rest-frame EW. Robust measurements of continuum flux are required to compute EWs. For objects with bright continuum (S/N $>5$) in the extracted 1D spectra, we derive the flux density in a clean window of $\pm150$~\AA\ near the emission line. For objects lacking confident continuum detection in our spectra, we derive the continuum flux from the best-fit SED models.

For $24$ sources in our $z=1.3-3.7$ sample (Table \ref{tab:opt_info}), we have detected at least one UV metal line (C~{\small IV}, O~{\small III}], C~{\small III}]) with S/N $>3$. The measured emission line fluxes and EWs are presented in Table \ref{tab:line_flux_z14} and Table \ref{tab:line_ew_z14}, respectively. In Figure \ref{fig:uv_spec} we show a subset of the sources with UV emission line detections. In most cases, the blended C~{\small III}]~$\lambda1908$ doublet is the strongest rest-frame UV emission line (other than Ly$\alpha$). We measured C~{\small III}] in $20$ extreme [O~{\small III}] emitters with S/N $>3$. For three sources in which the spectral region around C~{\small III}] is contaminated by sky line residuals, we nonetheless detect emission from O~{\small III}]~$\lambda1666$. In the remaining source we have a detection of the C~{\small IV} line and a tentative detection ($2.4\sigma$) of the C~{\small III}] emission.

For the $20$ galaxies in our sample with C~{\small III}] detections, the measured fluxes are from $3.1\times10^{-18}$~erg~s$^{-1}$~cm$^{-2}$ to $3.2\times10^{-17}$~erg~s$^{-1}$~cm$^{-2}$. We compute the C~{\small III}] EWs in the range from $1.6$~\AA\ to $18.7$~\AA, with a median value of $5.8$~\AA. In comparison, \citet{Du2020} report a similar median C~{\small III}] EW ($4.0$~\AA) for $33$ EELGs at $z\sim2$ which are selected following similar criteria in \citetalias{Tang2019}. Our C~{\small III}] EWs are also consistent with the values of $z\sim2$ EELGs in \citet{Stark2014} (median EW$_{\rm{CIII]}}=7.1$~\AA) and in \citet{Maseda2017} (median EW$_{\rm{CIII]}}=4.8$~\AA). On the other hand, investigations of more massive, typical star-forming galaxies usually find much lower C~{\small III}] EWs. The median C~{\small III}] EW of the individual C~{\small III}] detections in our sample is $\sim3-4\times$ larger than that of star-forming galaxies at $z\sim1$ (median EW$_{\rm{CIII]}}=1.30$~\AA; \citealt{Du2017}), $z\sim2-3$ (median EW$_{\rm{CIII]}}=2.0$~\AA; \citealt{LeFevre2019}), and $z\sim3$ (average EW$_{\rm{CIII]}}=1.67$~\AA\ derived from composite spectrum; \citealt{Shapley2003}). 

While the 20 C~{\small III}] emitters in our sample represents 
a significant boost to the statistics in this very large sSFR regime, 
the majority of galaxies we observed do not show C~{\small III}] emission. 
In many of these cases, the galaxies with non-detections are very faint in the 
continuum. As a result, our upper limits are often not  
sufficient to detect C~{\small III}] or other lines. Indeed, the median $3\sigma$ flux limit for galaxies without a C~{\small III}] detection is $6.6\times10^{-18}$~erg~s$^{-1}$~cm$^{-2}$. Given the range of continuum magnitudes, we 
calculate a median upper limit of C~{\small III}] EW =$14.8$~\AA\ for our sample of non-detected sources. Clearly 
many of these systems may 
hold large EW ($>5-10$~\AA) C~{\small III}] line emission below our detection limits. In the future, we aim to obtain 
deeper exposures which should allow us to 
uncover emission lines in many of these galaxies. For the 
analysis in this paper, we will consider the non-detections when 
assessing the range of line strengths seen in our sample.  

The second most commonly detected UV metal line in our sample 
is O~{\small III}]~$\lambda1666$. Seven galaxies 
are seen with O~{\small III}]~$\lambda1666$ above 
$>3\sigma$ (see Table \ref{tab:line_flux_z14} and Table \ref{tab:line_ew_z14}). The O~{\small III}]~$\lambda1666$ EWs range from $2.1$~\AA\ to $6.7$~\AA\ with a median value of $4.0$~\AA. This is larger than the average O~{\small III}]~$\lambda1666$ EW ($0.23$~\AA) of typical star-forming galaxies at $z\sim3$ in \citet{Shapley2003} but comparable to the EWs of 
O~{\small III}]~$\lambda1666$ detections in other EELGs at $z\sim2$ \citep[e.g.][]{Stark2014,Mainali2020,Du2020}. The O~{\small III}]~$\lambda1661$ emission line is 
found to be weaker, showing up in only one object, COSMOS-22402 (Figure \ref{fig:uv_spec}). The doublet flux ratio in this 
system is O~{\small III}] $1666:1661=1.65$. For other galaxies in our sample with O~{\small III}]~$\lambda1666$ detections, we can place a limit on the 
O~{\small III}] doublet ratio of $1666:1661>1.2$ at $3\sigma$. This is consistent with both the observed doublet ratios \citep[e.g.][]{Stark2014,Senchyna2017,Berg2019,Mainali2020} and the ratio computed from theoretical transition probabilities ($1666:1661\simeq2.5$; \citealt{FroeseFischer1985}). 

We also detected the Si~{\small III}]~$\lambda\lambda1883,1892$ emission lines in 
COSMOS-04064 (Figure \ref{fig:uv_spec}). The Si~{\small III}] EWs (Si~{\small III}]~$\lambda1883$ EW $=2.2\pm0.3$~\AA, Si~{\small III}]~$\lambda1892$ EW $=2.6\pm0.2$~\AA) of this object are comparable to those measured for low mass, metal-poor galaxies at $z\sim0-2$ \citep{Berg2018,Berg2019,Mainali2020}. Two systems show Mg~{\small II}~$\lambda\lambda2796,2803$ emission (see Figure \ref{fig:uv_spec}).  Both 
sources show rather intense line emission, with COSMOS-04064 
(Mg~{\small II}~$\lambda2796$ EW $=7.5\pm0.4$~\AA, Mg~{\small II}~$\lambda2803$ EW $=2.1\pm0.3$~\AA) 
appearing stronger than COSMOS-18358 (Mg~{\small II}~$\lambda2796$ EW $=2.9\pm0.2$~\AA, Mg~{\small II}~$\lambda2803$ EW $=2.2\pm0.2$~\AA).  Since Mg~{\small II} emission is resonantly scattered by neutral gas, it has been suggested that Mg~{\small II} line properties may provide a 
valuable probe of the escape fraction of Ly$\alpha$ photons 
\citep[e.g.][]{Henry2018,Feltre2018}. We will discuss the implications of the Mg~{\small II} strengths in a separate paper.

Nebular He~{\small II} emission has been seen in both nearby and high-redshift star-forming galaxies \citep[e.g.][]{Erb2010,Shirazi2012,Stark2014,Senchyna2017,Berg2018,Senchyna2019a,Senchyna2019b,Saxena2020}, reflecting a very hard ionizing spectrum. The origin of strong nebular He~{\small II}~$\lambda1640$ and $4686$ lines is still a matter of debate, with possibilities including shocks, X-ray binaries, or metal-poor massive stars (e.g., \citealt{Thuan2005,Senchyna2017}; see also \citealt{Shirazi2012}). These papers generally showed that observed He~{\small II} strengths typically appear larger 
than predicted by stellar population models, potentially suggesting that these 
models are missing sources of energetic photons at low metallicity. We detect confident He~{\small II}~$\lambda1640$ emission ($8.5\sigma$) in only one galaxy (COSMOS-11530) in our sample.  The spectrum of this galaxy also shows O~{\small III}]~$\lambda1666$ and C~{\small III}]~$\lambda1908$ (Figure \ref{fig:uv_spec}). The He~{\small II} EW is $2.2\pm0.3$~\AA, comparable to the EWs derived for metal-poor dwarf galaxies at $z\sim0-2$ \citep[e.g.][]{Erb2010,Senchyna2017,Senchyna2019b,Berg2019}. However, the full width at half maximum 
(FWHM) of He~{\small II} ($760$ km/s; uncorrected for spectral resolution) is nearly twice that of the O~{\small III}]~$\lambda1666$ emission line ($390$ km/s at $5100$~\AA, i.e., close to the spectral resolution). The broad profile suggests a significant contribution from stellar winds, as is commonly seen in star-forming galaxies.  
Higher spectral resolution is required to distinguish the nebular and stellar components of He~{\small II} in this system.

Metal poor galaxies have also been shown to power nebular C~{\small IV} emission \citep[e.g.][]{Berg2016,Berg2019,Senchyna2019b}. 
We detect the blended C~{\small IV}~$\lambda1549$ doublet with S/N $>3$ in  three galaxies  (COSMOS-22402, UDS-07665, and UDS-08735; Figure \ref{fig:uv_spec}). 
The data imply  large C~{\small IV} EWs in COSMOS-22402 ($4.7\pm0.9$~\AA) and UDS-07665 ($8.6\pm2.8$~\AA), with C~{\small IV}/C~{\small III}] ratios of $0.46-0.66$. For UDS-08735, we see even stronger C~{\small IV} (EW $=20.4\pm4.7$~\AA), with a very large C~{\small IV}/C~{\small III}] ratio ($2.8$).  We do not see any interstellar or stellar absorption in the vicinity of these lines, but we cannot rule out 
a modest level of emission line filling (and hence larger C~{\small IV} line strengths).  Higher resolution spectra will provide a more robust measure of the total C~{\small IV}  EWs in these systems. In Section \ref{sec:ionizing_source}, we will come back to investigate the ionizing nature of these three systems.

For one of the C~{\small III}] emitters in our sample (COSMOS-04064), we are not
able to constrain rest-wavelengths shorter than 
Si~{\small III}] due to the low spectral sensitivity at the blue end of IMACS spectra. This galaxy was previously observed with Keck/LRIS in \citet{Du2020}, with multiple UV lines detected 
(C~{\small IV}~$\lambda1549$, He~{\small II}~$\lambda1640$, O~{\small III}]~$\lambda1666$, 
C~{\small III}]~$\lambda1908$). The 
C~{\small III}] EW ($6.9\pm0.3$~\AA) measured from \citet{Du2020} is
consistent with the value measured from our IMACS spectrum ($7.2\pm0.3$~\AA).  When considering this galaxy in the 
following analysis, we will augment the fluxes and EWs 
measurements with the fluxes and EWs presented in \citet{Du2020}.


\begin{landscape}
\begin{table}
\centering
\begin{tabular}{|c|c|c|c|c|c|c|c|c|c|c|}
\hline
Target & R.A. & Decl. & $z_{\rm{spec}}$ & $m_{\rm{F814W}}$ & $M_{\rm{UV}}$ & UV slope & $E(B-V)$ & EW$_{\rm{[OIII]+H}\beta}$ (\AA) & $\log{(\rm{M}_{\star}/\rm{M}_{\odot})}$ & sSFR (Gyr$^{-1}$) \\
\hline 
\hline
COSMOS-04064 & 10:00:24.375 & +02:13:08.921 & $1.5019$ & $23.50\pm0.01$ & $-20.57$ & $-2.01$ & $0.12$ & $2192\pm530$ & $8.60^{+0.05}_{-0.07}$ & $61^{+11}_{-6}$ \\ 
COSMOS-04156 & 10:00:43.037 & +02:13:11.174 & $2.1883$ & $24.49\pm0.02$ & $-20.52$ & $-2.08$ & $0.06$ & $1389\pm258$ & $8.78^{+0.06}_{-0.04}$ & $45^{+6}_{-7}$ \\ 
COSMOS-04432 & 10:00:32.201 & +02:13:21.399 & $1.6206$ & $24.86\pm0.04$ & $-18.91$ & $-1.29$ & $0.33$ & $1608\pm139$ & $8.57^{+0.03}_{-0.03}$ & $614^{+239}_{-190}$ \\ 
COSMOS-04870 & 10:00:17.219 & +02:13:37.980 & $2.1023$ & $24.65\pm0.05$ & $-20.04$ & $-1.53$ & $0.16$ & $420\pm52$ & $9.21^{+0.08}_{-0.08}$ & $14^{+7}_{-4}$ \\ 
COSMOS-11530 & 10:00:28.638 & +02:17:48.674 & $2.0969$ & $24.08\pm0.01$ & $-20.93$ & $-2.31$ & $0.00$ & $1684\pm65$ & $8.19^{+0.01}_{-0.02}$ & $327^{+11}_{-14}$ \\ 
COSMOS-16680 & 10:00:48.029 & +02:20:57.824 & $3.1846$ & $24.00\pm0.02$ & $-21.63$ & $-1.99$ & $0.12$ & $1102\pm118$ & $8.94^{+0.14}_{-0.13}$ & $90^{+35}_{-27}$ \\ 
COSMOS-18358 & 10:00:40.111 & +02:22:00.462 & $1.6486$ & $23.09\pm0.02$ & $-21.33$ & $-2.15$ & $0.12$ & $639\pm37$ & $9.51^{+0.02}_{-0.03}$ & $12^{+1}_{-1}$ \\ 
COSMOS-22402 & 10:00:17.831 & +02:24:26.350 & $2.2751$ & $24.62\pm0.05$ & $-20.35$ & $-1.79$ & $0.18$ & $682\pm55$ & $8.59^{+0.06}_{-0.06}$ & $107^{+14}_{-13}$ \\ 
COSMOS-24660 & 10:00:34.285 & +02:25:58.495 & $1.5897$ & $24.84\pm0.04$ & $-19.39$ & $-1.87$ & $0.15$ & $1050\pm79$ & $8.41^{+0.09}_{-0.08}$ & $43^{+10}_{-9}$ \\ 
UDS-07447 & 02:17:18.162 & -05:15:06.275 & $1.5972$ & $24.46\pm0.02$ & $-19.60$ & $-1.61$ & $0.12$ & $983\pm48$ & $8.64^{+0.05}_{-0.05}$ & $25^{+4}_{-3}$ \\ 
UDS-07665 & 02:17:33.781 & -05:15:02.848 & $2.2955$ & $25.45\pm0.06$ & $-19.35$ & $-2.18$ & $0.11$ & $1800\pm127$ & $8.11^{+0.09}_{-0.08}$ & $252^{+87}_{-86}$ \\ 
UDS-08735 & 02:17:34.564 & -05:14:48.779 & $2.2939$ & $25.29\pm0.05$ & $-18.98$ & $-0.67$ & $0.44$ & $460\pm24$ & $10.84^{+0.08}_{-0.09}$ & $1^{+0}_{-1}$ \\ 
UDS-11010 & 02:17:14.707 & -05:14:20.245 & $1.6637$ & $26.03\pm0.08$ & $-19.03$ & $-2.83$ & $0.00$ & $1363\pm28$ & $7.62^{+0.06}_{-0.06}$ & $79^{+10}_{-9}$ \\ 
UDS-11457 & 02:17:08.085 & -05:14:16.134 & $2.1821$ & $25.04\pm0.04$ & $-19.89$ & $-1.87$ & $0.12$ & $1572\pm183$ & $8.55^{+0.09}_{-0.08}$ & $148^{+26}_{-36}$ \\ 
UDS-11693 & 02:17:03.893 & -05:14:13.664 & $2.1854$ & $24.55\pm0.02$ & $-20.38$ & $-1.82$ & $0.14$ & $869\pm45$ & $8.85^{+0.09}_{-0.09}$ & $35^{+10}_{-8}$ \\ 
UDS-12154 & 02:17:52.098 & -05:14:09.985 & $2.3065$ & $24.85\pm0.04$ & $-20.08$ & $-1.50$ & $0.23$ & $689\pm134$ & $9.44^{+0.09}_{-0.08}$ & $6^{+2}_{-2}$ \\ 
UDS-12539 & 02:17:53.733 & -05:14:03.196 & $1.6211$ & $24.30\pm0.02$ & $-19.92$ & $-1.97$ & $0.12$ & $1377\pm44$ & $8.33^{+0.09}_{-0.06}$ & $123^{+15}_{-22}$ \\ 
UDS-19167 & 02:17:43.535 & -05:12:43.610 & $2.1833$ & $24.95\pm0.03$ & $-20.27$ & $-2.41$ & $0.05$ & $2335\pm178$ & $7.95^{+0.02}_{-0.02}$ & $232^{+13}_{-12}$ \\ 
UDS-21196 & 02:17:33.633 & -05:12:17.791 & $2.1585$ & $24.13\pm0.02$ & $-21.08$ & $-2.54$ & $0.01$ & $343\pm23$ & $9.17^{+0.04}_{-0.04}$ & $6^{+1}_{-1}$ \\ 
UDS-27151 & 02:17:36.141 & -05:11:06.180 & $2.1539$ & $24.33\pm0.03$ & $-20.68$ & $-2.15$ & $0.10$ & $946\pm72$ & $8.80^{+0.10}_{-0.07}$ & $39^{+9}_{-10}$ \\ 
UDS-29267 & 02:17:25.322 & -05:10:40.397 & $1.5190$ & $23.85\pm0.01$ & $-20.71$ & $-2.53$ & $0.00$ & $2787\pm80$ & $8.16^{+0.01}_{-0.01}$ & $316^{+4}_{-3}$ \\ 
UDS-30015 & 02:17:36.517 & -05:10:31.256 & $1.6649$ & $24.70\pm0.03$ & $-19.81$ & $-2.06$ & $0.01$ & $1332\pm73$ & $7.86^{+0.02}_{-0.02}$ & $105^{+12}_{-15}$ \\ 
UDS-30274 & 02:17:21.117 & -05:10:28.812 & $1.4570$ & $23.76\pm0.02$ & $-20.20$ & $-1.83$ & $0.09$ & $321\pm28$ & $9.29^{+0.05}_{-0.06}$ & $6^{+1}_{-1}$ \\ 
UDS-31649 & 02:17:06.433 & -05:10:13.584 & $1.4589$ & $23.72\pm0.01$ & $-20.54$ & $-2.26$ & $0.08$ & $1014\pm85$ & $8.96^{+0.04}_{-0.04}$ & $12^{+1}_{-1}$ \\ 
\hline
\end{tabular}
\caption{Coordinates, spectroscopic redshifts, F814W magnitudes, absolute UV magnitudes, UV slopes, dust extinction, [O~{\scriptsize III}]+H$\beta$ EWs, stellar masses, and specific star formation rates of the $24$ galaxies at $z=1.3-3.7$ with rest-frame UV metal emission line detections (C~{\scriptsize IV}, O~{\scriptsize III}], or C~{\scriptsize III}] detected with S/N $>3$). Redshifts ($z_{\rm{spec}}$) are derived from [O~{\scriptsize III}]~$\lambda5007$ emission lines, except for COSMOS-16680 whose redshift is derived based on the O~{\scriptsize III}]~$\lambda1666$ emission line. Dust extinction, stellar masses, and specific star formation rates are derived from BEAGLE modeling (Section \ref{sec:modeling}).}
\label{tab:opt_info}
\end{table}
\end{landscape}


\begin{table*}
\begin{tabular}{|c|c|c|c|c|c|}
\hline
Target & $z_{\rm{spec}}$ & C~{\small IV}~$\lambda1549$ & O~{\small III}]~$\lambda1661$ & O~{\small III}]~$\lambda1666$ & C~{\small III}]~$\lambda1908$ \\
 & & ($\times10^{-18}$~erg~s$^{-1}$~cm$^{-2}$) & ($\times10^{-18}$~erg~s$^{-1}$~cm$^{-2}$) & ($\times10^{-18}$~erg~s$^{-1}$~cm$^{-2}$) & ($\times10^{-18}$~erg~s$^{-1}$~cm$^{-2}$) \\
\hline 
\hline
COSMOS-04064 & $1.5019$ & ... & ... & ... & $29.99\pm1.24$ \\ 
COSMOS-04156 & $2.1883$ & $<5.39$ & $<1.90$ & $<1.87$ & $8.87\pm0.89$ \\ 
COSMOS-04432 & $1.6206$ & ... & $<4.15$ & $<5.28$ & $17.07\pm1.44$ \\ 
COSMOS-04870 & $2.1023$ & $<5.27$ & $<1.59$ & $<1.52$ & $3.15\pm0.37$ \\ 
COSMOS-11530 & $2.0969$ & $<6.04$ & $<8.64$ & $7.58\pm0.89$ & $16.82\pm1.02$ \\ 
COSMOS-16680 & $3.1846$ & ... & ... & $4.28\pm0.93$ & ... \\ 
COSMOS-18358 & $1.6486$ & ... & $<4.54$ & $<4.31$ & $18.62\pm1.09$ \\ 
COSMOS-22402 & $2.2751$ & $8.92\pm1.61$ & $4.26\pm1.18$ & $7.04\pm0.96$ & $19.17\pm1.17$ \\ 
COSMOS-24660 & $1.5897$ & ... & $<5.07$ & $<5.56$ & $6.65\pm1.61$ \\ 
UDS-07447 & $1.5972$ & $<18.92$ & $<5.99$ & $<7.21$ & $10.43\pm2.35$ \\ 
UDS-07665 & $2.2955$ & $6.19\pm2.05$ & $<3.95$ & $4.76\pm1.15$ & $9.32\pm2.92$ \\ 
UDS-08735 & $2.2939$ & $25.13\pm5.74$ & $<3.98$ & $<5.32$ & $9.06\pm3.83$ \\ 
UDS-11010 & $1.6637$ & ... & $<19.79$ & $<18.77$ & $7.12\pm1.34$ \\ 
UDS-11457 & $2.1821$ & $<16.36$ & $<5.56$ & $4.44\pm1.63$ & $5.40\pm1.50$ \\ 
UDS-11693 & $2.1854$ & $<14.73$ & $<6.24$ & $<6.14$ & $5.59\pm1.56$ \\ 
UDS-12154 & $2.3065$ & $<12.96$ & ... & $5.24\pm1.44$ & ... \\ 
UDS-12539 & $1.6211$ & ... & $<23.77$ & $<22.72$ & $10.12\pm2.19$ \\ 
UDS-19167 & $2.1833$ & $8.97\pm4.50$ & $<6.54$ & $6.95\pm2.11$ & ... \\ 
UDS-21196 & $2.1585$ & $<17.54$ & $<6.00$ & $<4.55$ & $6.37\pm1.48$ \\ 
UDS-27151 & $2.1539$ & $<18.51$ & $<5.11$ & $<7.01$ & $9.93\pm2.68$ \\ 
UDS-29267 & $1.5190$ & ... & $<7.30$ & $<5.09$ & $31.56\pm3.72$ \\ 
UDS-30015 & $1.6649$ & ... & $<15.31$ & $<19.36$ & $8.34\pm1.79$ \\ 
UDS-30274 & $1.4570$ & ... & ... & ... & $14.03\pm3.17$ \\ 
UDS-31649 & $1.4589$ & ... & ... & ... & $6.86\pm2.15$ \\ 
\hline
\end{tabular}
\caption{Rest-frame UV metal line fluxes of the $24$ sources with one or more UV metal line detections. We provide the $3\sigma$ upper limits for non-detections.}
\label{tab:line_flux_z14}
\end{table*}


\begin{table*}
\begin{tabular}{|c|c|c|c|c|c|}
\hline
Target & $z_{\rm{spec}}$ & C~{\small IV}~$\lambda1549$ & O~{\small III}]~$\lambda1661$ & O~{\small III}]~$\lambda1666$ & C~{\small III}]~$\lambda1908$ \\
 & & (\AA) & (\AA) & (\AA) & (\AA) \\
\hline 
\hline
COSMOS-04064 & $1.5019$ & $1.5\pm0.2^{\rm{a}}$ & ... & $1.5\pm0.3^{\rm{a}}$ & $7.19\pm0.30$ \\ 
COSMOS-04156 & $2.1883$ & $<2.25$ & $<0.91$ & $<0.90$ & $5.60\pm0.56$ \\ 
COSMOS-04432 & $1.6206$ & ... & $<3.87$ & $<4.93$ & $18.72\pm1.58$ \\ 
COSMOS-04870 & $2.1023$ & $<3.03$ & $<1.00$ & $<0.96$ & $2.50\pm0.30$ \\ 
COSMOS-11530 & $2.0969$ & $<1.55$ & $<2.61$ & $2.29\pm0.27$ & $7.01\pm0.43$ \\ 
COSMOS-16680 & $3.1846$ & ... & ... & $2.10\pm0.46$ & ... \\ 
COSMOS-18358 & $1.6486$ & ... & $<0.52$ & $<0.50$ & $2.82\pm0.16$ \\ 
COSMOS-22402 & $2.2751$ & $4.68\pm0.85$ & $2.57\pm0.71$ & $4.24\pm0.58$ & $14.54\pm0.89$ \\ 
COSMOS-24660 & $1.5897$ & ... & $<3.06$ & $<3.35$ & $5.17\pm1.25$ \\ 
UDS-07447 & $1.5972$ & $<8.65$ & $<3.07$ & $<3.70$ & $6.58\pm1.49$ \\ 
UDS-07665 & $2.2955$ & $8.55\pm2.83$ & $<3.68$ & $6.73\pm1.63$ & $17.75\pm5.56$ \\ 
UDS-08735 & $2.2939$ & $20.43\pm4.66$ & $<3.16$ & $<4.22$ & $6.71\pm2.84$ \\ 
UDS-11010 & $1.6637$ & ... & $<20.72$ & $<19.66$ & $11.55\pm2.18$ \\ 
UDS-11457 & $2.1821$ & $<11.32$ & $<4.41$ & $3.53\pm1.29$ & $5.78\pm1.61$ \\ 
UDS-11693 & $2.1854$ & $<8.67$ & $<4.03$ & $<3.96$ & $4.44\pm1.24$ \\ 
UDS-12154 & $2.3065$ & $<8.97$ & ... & $3.97\pm1.10$ & ... \\ 
UDS-12539 & $1.6211$ & ... & $<11.64$ & $<11.13$ & $5.82\pm1.26$ \\ 
UDS-19167 & $2.1833$ & $5.04\pm2.53$ & $<4.32$ & $4.59\pm1.39$ & ... \\ 
UDS-21196 & $2.1585$ & $<4.21$ & $<1.73$ & $<1.31$ & $2.60\pm0.60$ \\ 
UDS-27151 & $2.1539$ & $<6.27$ & $<2.03$ & $<2.79$ & $5.33\pm1.44$ \\ 
UDS-29267 & $1.5190$ & ... & $<1.49$ & $<1.04$ & $8.49\pm1.00$ \\ 
UDS-30015 & $1.6649$ & ... & $<8.41$ & $<10.64$ & $5.77\pm1.24$ \\ 
UDS-30274 & $1.4570$ & ... & ... & ... & $4.14\pm0.94$ \\ 
UDS-31649 & $1.4589$ & ... & ... & ... & $1.64\pm0.52$ \\ 
\hline
\end{tabular}
\caption{Equivalent widths of the rest-frame UV metal emission lines. We provide the $3\sigma$ upper limits for non-detections. {\bf Note.} $^{\rm{a}}$From \citet{Du2020}.}
\label{tab:line_ew_z14}
\end{table*}

\subsection{Rest-frame UV spectra at $z\sim4-6$} \label{sec:spec_z46}


\begin{table*}
\begin{tabular}{|c|c|c|c|c|}
\hline
Target & Line & $\lambda_{\rm{rest}}$ & Line Flux & EW$_0$ \\
 & & (\AA) & ($\times10^{-18}$~erg~s$^{-1}$~cm$^{-2}$) & (\AA) \\
\hline 
\hline
COSMOS-18502 & [C~{\small III}]~$\lambda1907$ & $1906.68$ & $<3.1$ ($5\sigma$) & $<4.0$ ($5\sigma$) \\
 & C~{\small III}]~$\lambda1909$ & $1908.73$ & $<3.1$ ($5\sigma$) & $<4.0$ ($5\sigma$) \\
COSMOS-19732 & [C~{\small III}]~$\lambda1907$ & $1906.68$ & $<3.0$ ($5\sigma$) & $<2.5$ ($5\sigma$) \\
 & C~{\small III}]~$\lambda1909$ & $1908.73$ & $<3.0$ ($5\sigma$) & $<2.5$ ($5\sigma$) \\
COSMOS-11116 & O~{\small III}]~$\lambda1661$ & $1660.81$ & $<1.8$ ($3\sigma$) & $<5.1$ ($3\sigma$) \\
 & O~{\small III}]~$\lambda1666$ & $1666.15$ & $3.1\pm0.5$ & $8.8\pm1.7$ \\
COSMOS-15365 & O~{\small III}]~$\lambda1661$ & $1660.81$ & $<3.4$ ($5\sigma$) & $<5.2$ ($5\sigma$) \\
 & O~{\small III}]~$\lambda1666$ & $1666.15$ & $<3.4$ ($5\sigma$) & $<5.2$ ($5\sigma$) \\
COSMOS-20187 & O~{\small III}]~$\lambda1661$ & $1660.81$ & $<4.3$ ($5\sigma$) & $<22$ ($5\sigma$) \\
 & O~{\small III}]~$\lambda1666$ & $1666.15$ & $<4.3$ ($5\sigma$) & $<22$ ($5\sigma$) \\
GOODS-S-36712 & [C~{\small III}]~$\lambda1907$ & $1906.68$ & $<2.4$ ($5\sigma$) & $<6.1$ ($5\sigma$) \\
 & C~{\small III}]~$\lambda1909$ & $1908.73$ & $<2.4$ ($5\sigma$) & $<6.1$ ($5\sigma$) \\
GOODS-S-38450 & [C~{\small III}]~$\lambda1907$ & $1906.68$ & $<2.2$ ($5\sigma$) & $<4.9$ ($5\sigma$) \\
 & C~{\small III}]~$\lambda1909$ & $1908.73$ & $<2.2$ ($5\sigma$) & $<4.9$ ($5\sigma$) \\
GOODS-S-39157 & [C~{\small III}]~$\lambda1907$ & $1906.68$ & $<2.5$ ($5\sigma$) & $<6.2$ ($5\sigma$) \\
 & C~{\small III}]~$\lambda1909$ & $1908.73$ & $<2.5$ ($5\sigma$) & $<6.2$ ($5\sigma$) \\
GOODS-S-40887 & [C~{\small III}]~$\lambda1907$ & $1906.68$ & $<2.6$ ($5\sigma$) & $<9.1$ ($5\sigma$) \\
 & C~{\small III}]~$\lambda1909$ & $1908.73$ & $<2.6$ ($5\sigma$) & $<9.1$ ($5\sigma$) \\
GOODS-S-41253 & [C~{\small III}]~$\lambda1907$ & $1906.68$ & $<2.4$ ($5\sigma$) & $<11$ ($5\sigma$) \\
 & C~{\small III}]~$\lambda1909$ & $1908.73$ & $<2.4$ ($5\sigma$) & $<11$ ($5\sigma$) \\
GOODS-S-46692 & [C~{\small III}]~$\lambda1907$ & $1906.68$ & $<1.9$ ($3\sigma$) & $<5.5$ ($3\sigma$) \\
 & C~{\small III}]~$\lambda1909$ & $1908.73$ & $<1.9$ ($3\sigma$) & $<5.5$ ($3\sigma$) \\
\hline
\end{tabular}
\caption{Rest-frame UV metal line constraints on the sources at $z\sim4-6$. EWs are given in rest-frame. For galaxies with spectroscopic redshift measurements, i.e., COSMOS-11116 and GOODS-S-46692, the upper limits on line fluxes and EWs are quoted as $3\sigma$. For objects with photometric redshift measurements, the upper limits are quoted as $5\sigma$.}
\label{tab:line_z46}
\end{table*}


\begin{figure}
\begin{center}
\includegraphics[width=0.9\linewidth]{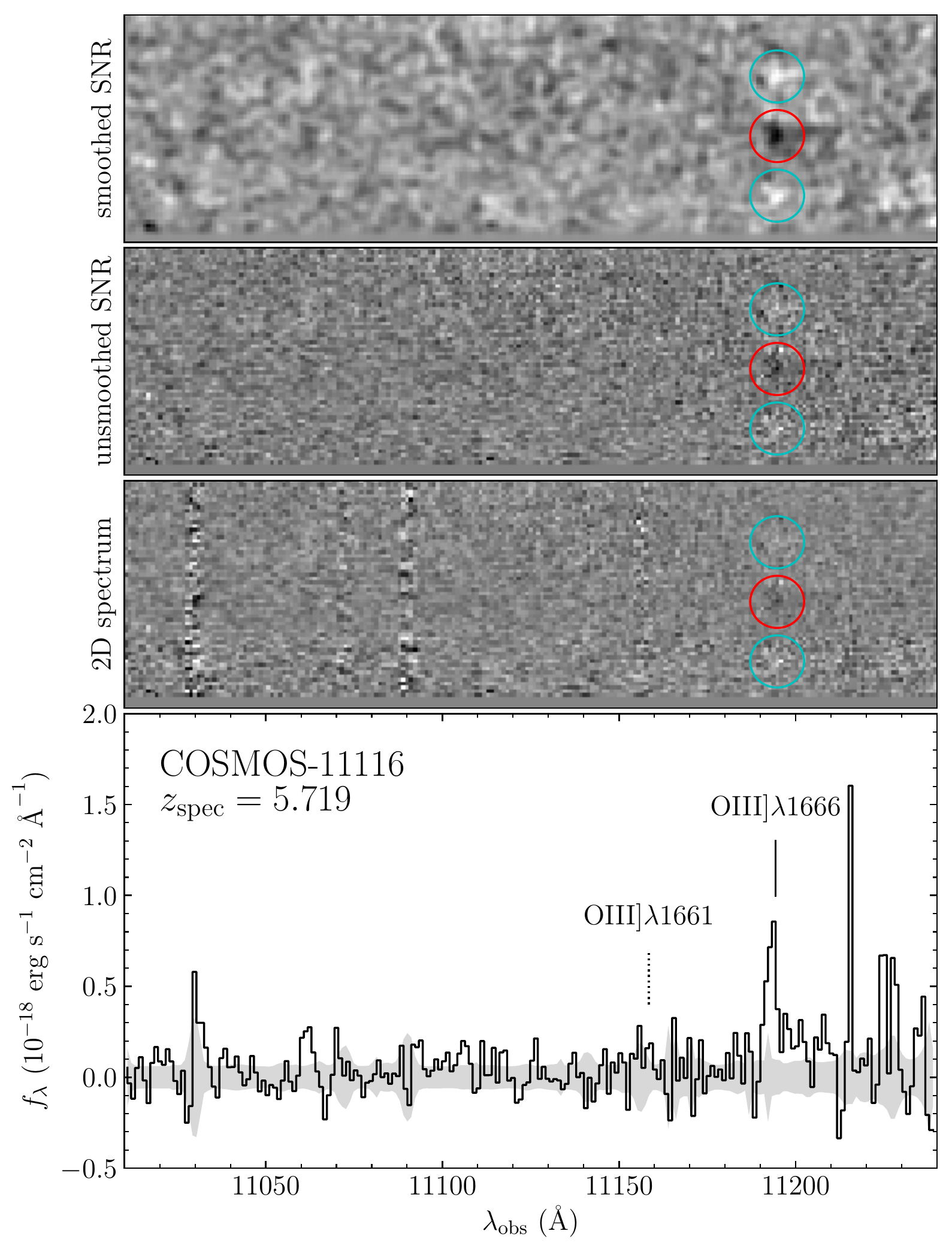}
\caption{Keck/MOSFIRE $Y$-band spectrum of COSMOS-11116. We identified an emission line feature at $11194.3$~\AA. This emission line is likely to be O~{\scriptsize III}]~$\lambda1666$ at $z=5.719$. The top panel shows the smoothed and unsmoothed 2D S/N map, as well as the 2D spectrum. Red circles show the positive peak of the emission line, while cyan circles show the negative peaks created in dithering. The bottom panel presents the flux-calibrated 1D spectrum, with $1\sigma$ error shown in grey. The detected emission line (O~{\scriptsize III}]~$\lambda1666$ at $z=5.719$) is marked by the solid black line, and the position where O~{\scriptsize III}]~$\lambda1661$ (undetected) should be is marked by the dotted black line.}
\label{fig:COS11116_spec}
\end{center}
\end{figure}

The near-infrared spectra of the eleven $z\sim4-6$ targets described in 
Section \ref{sec:obs_z46} allow constraints to be placed on the strengths 
of rest-frame UV metal lines (O~{\small III}], C~{\small III}]). 
Each MOSFIRE $Y$-band spectrum is mostly ($\sim87\%$) free from strong OH lines.
In all but one case, the spectra do not show emission lines, 
so we put upper limits on the UV metal lines. 
For sources only having photometric redshifts, we report 
$5\sigma$ upper limits. In the one case where the spectroscopic redshift is 
already known (see \S \ref{sec:obs_z46}), we 
can more confidently identify faint features because of the narrow  
wavelength window where the line is expected.  For this object, we thus present $3\sigma$ upper limits on non-detected lines.
To compute constraints on the UV line EWs, we estimate the underlying broadband 
continuum using our best-fit SED models (Section \ref{sec:modeling}).
For sources in the redshift range $3.8<z<5.0$, the IRAC color is sensitive to the 
H$\alpha$ strength. We quote H$\alpha$ EWs implied by our photoionization model 
fits. We summarize the results for each source below and in Table \ref{tab:line_z46}. 

The GOODS-S mask has six $z>4$ objects (see Table \ref{tab:nir_obs}).
GOODS-S-46692 is the only galaxy in our sample with a spectroscopic redshift from 
earlier efforts \citep{Vanzella2008}, with weak Ly$\alpha$ emission 
(Ly$\alpha$ EW $=7.7\pm1.8$~\AA) indicating a redshift of $z_{\rm{Ly}\alpha}=4.811$.
The IRAC color is blue, suggesting the likely presence of strong H$\alpha$ 
emission. The BEAGLE SED modeling suggests the IRAC color can be reproduced with H$\alpha$ EW $=693\pm86$~\AA. 
In this redshift range, the Y-band spectrum 
covers the C~{\small III}] doublet. Assuming that the velocity offset of Ly$\alpha$ with respect to C~{\small III}] is between $-200$ and $1000$ km s$^{-1}$ \citep[e.g.][]{Erb2014,Stark2017}, the [C~{\small III}]~$\lambda1907$ emission line will be located between $11043.5$ and $11087.3$~\AA, and the C~{\small III}]~$\lambda1909$ line will be located between $11055.4$ and $11099.3$~\AA.  
No emission lines are seen in either wavelength range. We note that two OH skylines are located in this wavelength window. However, the wavelength separation of individual emission lines of C~{\small III}] at this redshift ($11.9$~\AA) guarantees that at least one of the two emission lines should be situated in a clean region of the spectrum. The $3\sigma$ upper limit of line flux for each individual component estimated from the clean regions is $1.9\times10^{-18}$~erg~s$^{-1}$~cm$^{-2}$.  Here 
and below, the flux limit is computed by summing the error spectrum in quadrature over $\sim200$ km/s ($8$~\AA), as discussed in Section \ref{sec:spec_z14}
To calculate the upper bound on the C~{\small III}] EW, we 
compute the underlying continuum flux using the best-fitting SED model.
Taken together with the flux limit, this implies a $3\sigma$ upper limit of 
rest-frame C~{\small III}] EW $=5.5$~\AA\ for each  component of the doublet.  
The measured Ly$\alpha$ line flux of GOODS-S-46692 is $7.43\times10^{-18}$~erg~s$^{-1}$~cm$^{-2}$, and the non-detection of C~{\small III}] indicates that the flux of doublet is less than $51\%$ of the observed Ly$\alpha$ flux, consistent with the C~{\small III}]/Ly$\alpha$ ratios observed in $z\simeq2$ young dwarf galaxies \citep{Stark2014,Du2020} and $z>6$ galaxies 
\citep{Stark2015a,Stark2017}. 

GOODS-S-41253 ($z_{\rm{phot}}=4.28$) has the bluest IRAC [3.6]-[4.5] color on the GOODS-S mask, consistent with a large H$\alpha$ EW ($829\pm182$~\AA). Given its photometric redshift, we expect the C~{\small III}]~$\lambda\lambda1907,1909$ emission line to lie in the $Y$-band.  No significant feature (S/N $>5$) is seen in the spectrum.
Using the clean regions between OH skylines, we estimated the $5\sigma$ upper limits of line flux and EW for GOODS-S-41253 are $2.4\times10^{-18}$~erg~s$^{-1}$~cm$^{-2}$ and $11$~\AA. These limits correspond to individual doublet components.  

The remaining four objects on the GOODS-S mask are GOODS-S-36712, GOODS-S-38450, GOODS-S-39157, 
and GOODS-S-40887. These galaxies have smaller flux excesses in the $[3.6]$ filter than the two sources
described above, suggesting weaker 
H$\alpha$ emission  (EW$=200-350$~\AA). No $>5\sigma$ feature was found in the 
MOSFIRE spectra, allowing us to place an upper limit on the C~{\small III}] 
line strength. Using the clean regions between OH skylines, we estimated the $5\sigma$ upper limits of line flux are $2.2-2.6\times10^{-18}$~erg~s$^{-1}$~cm$^{-2}$, corresponding to $5\sigma$ EW upper limits of $4.9-11$~\AA\ for individual 
components of the doublet. 

The COSMOS mask contains five $z\gtrsim4$ galaxies, two of which 
(COSMOS-18502 and COSMOS-19732)  
have IRAC colors indicative of strong H$\alpha$ emission ($400\pm24$~\AA\ and $1000\pm101$~\AA, 
respectively). Visual inspection of the MOSFIRE spectra reveals no convincing line features.
The $5\sigma$ upper limit on the line flux between OH skylines is $3.1\times10^{-18}$~erg~s$^{-1}$~cm$^{-2}$ for COSMOS-18502 and $3.0\times10^{-18}$~erg~s$^{-1}$~cm$^{-2}$ for COSMOS-19732. Using the continuum from the 
best-fit SED models, we compute a $5\sigma$ EW upper limit (for individual 
doublet components) of $4.0$~\AA\ for COSMOS-18502 and $2.5$~\AA\ for COSMOS-19732.

The COSMOS mask additionally contains three objects with photometric redshifts in 
the range $z=5.6-5.8$, allowing us to constrain the strength of OIII] emission. 
The first of these sources, COSMOS-11116, has a photometric redshift of $z_{\rm{phot}}=5.62$. 
We detect a $6.2\sigma$ emission line feature centered at $11194.3$~\AA\ with flux of
$3.1\pm0.5\times10^{-18}$~erg~s$^{-1}$~cm$^{-2}$ (Figure \ref{fig:COS11116_spec}), which is at the
expected spatial position of COSMOS-11116.  The emission line is unresolved, with FWHM similar to the
spectral resolution ($3.3$~\AA). It is shown that the emission line has the standard
negative-positive-negative pattern resulting from the subtraction of AB dither pattern, indicating that
the emission line is present in both dither positions. Given the photometric redshift, we 
conclude that the line is most likely either 
O~{\small III}]~$\lambda1666$ or O~{\small III}]~$\lambda1661$.
The separation between O~{\small III}] doublets is $\simeq35$~\AA\ at the expected redshift, which is more than $10\times$ the FWHM of the emission line, indicating that the O~{\small III}] doublets must be resolved in the spectrum. If the emission line feature is O~{\small III}]~$\lambda1666$ (O~{\small III}]~$\lambda1661$), the systematic redshift of COSMOS-11116 would be $z=5.719$ ($z=5.740$), which is consistent with the photometric redshift. Using the spectroscopic redshift derived from O~{\small III}]~$\lambda1666$ (O~{\small III}]~$\lambda1661$), we search for O~{\small III}]~$\lambda1661$ (O~{\small III}]~$\lambda1666$) which should be located at $11158.4$~($11230.3$)~\AA. No convincing ($>5\sigma$) emission line feature is detected at the expected position.  Theoretical transition probabilities imply a O~{\small III}]~$\lambda1666$/O~{\small III}]~$\lambda1661$ ratio $=2.5$ \citep{FroeseFischer1985}, and the observed doublet ratio found in high-redshift star-forming galaxies is $1666:1661>1$ \citep[e.g.][]{Stark2014,Mainali2020}. Therefore, we expect the O~{\small III}]~$\lambda1666$ emission line to be stronger than O~{\small III}]~$\lambda1661$. As a result, we tentatively identify this emission line as O~{\small III}]~$\lambda1666$, implying a spectroscopic redshift of $z=5.719$. Using the far-UV continuum flux density derived from the best-fitting BEAGLE model, we compute the rest-frame O~{\small III}]~$\lambda1666$ EW ($=8.8\pm1.7$~\AA). For the O~{\small III}]~$\lambda1661$ component, we estimate a $3\sigma$ upper limit of $=1.8\times10^{-18}$~erg~s$^{-1}$~cm$^{-2}$. This implies a O~{\small III}] doublet ratio of $1666:1661>1.7$ at $3\sigma$, which is consistent with both the theoretical and observed doublet ratio. The corresponding $3\sigma$ upper limit of O~{\small III}]~$\lambda1661$ EW is $5.1$~\AA.

The other two objects we targeted at $z=5.6-5.8$ are COSMOS-15365 ($z_{\rm{phot}}=5.64$) and COSMOS-20187 ($z_{\rm{phot}}=5.73$).  No significant ($>5\sigma$) feature is detected in either 
spectrum. Using the clean regions between OH skylines, we estimate upper limits for the flux and EW of individual components of the O~{\small III}] doublets.  For COSMOS-15365, we derive 5$\sigma$ 
upper limits of $3.4\times10^{-18}$~erg~s$^{-1}$~cm$^{-2}$ and $5.2$~\AA; whereas for COSMOS-20187, we 
compute $4.3\times10^{-18}$~erg~s$^{-1}$~cm$^{-2}$ for the flux limit, and $22$~\AA\ for the EW limit. 

\section{The Physical Nature of Galaxies with Intense Rest-Frame UV Nebular Emission} \label{sec:phy}

We have presented detection of rest-frame UV metal lines in $24$ galaxies at $z\simeq 1.3-3.7$ selected to have rest-frame optical line properties similar to those in the reionization era. The emission lines detected in our survey include C~{\small III}], O~{\small III}], C~{\small IV}, Mg~{\small II}, Si~{\small III}], and  He~{\small II}, with equivalent widths reaching close to the values seen in sources at $z>6$.  
In this section, we briefly summarize the physical properties of the UV line emitters in our sample (Section \ref{sec:properties}) and then investigate the nature of the ionizing sources responsible for powering the lines, considering whether any of our targets 
require AGN to explain the observed spectra (Section \ref{sec:ionizing_source}).

\subsection{Gas Conditions and Stellar Populations} \label{sec:properties}

Previous studies have demonstrated that rest-frame UV lines tend to be very 
prominent in metal poor low mass galaxies that are in the midst 
of a significant star formation episode \citep[e.g.][]{Stark2014,Feltre2016,Gutkin2016,Jaskot2016,Senchyna2017,Senchyna2019b,Plat2019,Mainali2020}. 
The BEAGLE photoionization models (Section \ref{sec:modeling}) suggest a similar picture for our 
targets. A summary of inferred properties is listed in Table \ref{tab:opt_info}. 
We have previously described the physical properties of 
extreme [O~{\small III}] emitters in our 
earlier papers (\citetalias{Tang2019}, Tang et al. in prep).
Here we briefly describe the average properties of the 
$24$ galaxies with rest-frame UV line detections below but note 
that the galaxies are very similar to those from the parent 
sample. Readers are directed to our earlier papers for more 
detailed discussion of the implied gas and stellar population properties. 

In nearly all of the $24$ galaxies with rest-frame UV metal line detections (more on the few exceptions in Section \ref{sec:ionizing_source}), models powered by stars are able to reproduce both broadband data and the observed nebular line detections (see fits in Figure \ref{fig:sed} for examples). The data are best-matched by models with low stellar masses (median values of $4.4\times10^8$ M$_{\odot}$) with gas that has moderately low metallicity and large ionization parameter (medians of $Z=0.2\ Z_\odot$ and $\log{U}=-1.7$).
As the UV continuum slopes tend to be very blue (median of $\beta=-2.0$), the best-fit solutions tend to have minimal reddening from dust (median value of $E(B-V)=0.12$). The rest-frame optical lines are very 
intense by virtue of our selection (median of EW$_{\rm{[OIII]+H\beta}}=1080$~\AA), requiring models with large 
specific star formation rates (median of $46$~Gyr$^{-1}$) and dominant young stellar populations.  

The rest-frame UV emission line spectra (in particular, C~{\small III}] and C~{\small IV}) are 
also sensitive to the C/O ratio of the ionized gas. Previous studies 
have suggested that for stellar populations to power the most intense 
C~{\small III}] emission lines (EW$_{\rm{CIII]}}>20$~\AA) seen at high redshift, 
the nebular gas must be enhanced in carbon, with super-solar 
C/O ratios \citep{Nakajima2018}. While none of our 
systems are detected above this C~{\small III}] EW threshold, several 
are very close. We nonetheless   
find that the BEAGLE fits prefer sub-solar C/O abundance ratios ($0.27$ or $0.52$ C/O$_{\odot}$), 
as the observed C~{\small III}] flux tends to be 
overpredicted in cases where solar C/O ratios are adopted.  
We can directly test this inference for systems with confident 
detections of both  C~{\small III}] and 
O~{\small III}].  With our current dataset, this limits us 
to COSMOS-11530 and COSMOS-22402, each of which have 
high quality (S/N $>7$) measurements of both lines.  
To infer the C/O ratios, we follow a similar procedure to other recent analyses \citep[e.g.][]{Erb2010,Stark2014,Mainali2020}. 
We first compute the ratio of doubly ionized 
carbon and oxygen with {\small PyNeb} \citep{Luridiana2015}, using the flux ratios of 
C~{\small III}] and O~{\small III}] together with an 
estimate of the electron temperature derived from the dust-corrected 
O~{\small III}]~$\lambda1666$/[O~{\small III}]~$\lambda5007$ flux ratio. We then apply a small ionization correction 
factor (ICF) to account for the fact that the fraction of carbon that is in the doubly-ionized state is not necessarily the same as 
that of oxygen. Following \citet{Mainali2020}, we estimate the 
ICFs using scaling relations derived in \citet{Berg2019}.  
If we conservatively adopt a metallicity 
range between $0.05\ Z_\odot$ and $0.2\ Z_\odot$, we find ICF values of $1.0$ for COSMOS-11530 and $1.4$ for COSMOS-22402. Applying these values, we compute C/O ratios of $\log$ C/O $=-0.95\pm0.11$ ($0.21\pm0.05$ C/O$_{\odot}$) 
for COSMOS-11530 and $\log$ C/O $=-0.48\pm0.13$ ($0.60\pm0.18$ C/O$_{\odot}$) for COSMOS-22402. Both values are sub-solar, 
consistent with the range implied by the BEAGLE models.

The spectra and imaging of our targets thus support a picture whereby 
most of the galaxies in our sample are moderately low mass and 
metal poor systems that have recently experienced a major star 
formation event. These results are very similar to 
what has been found in the literature for similar galaxies 
\citep[e.g.][]{Erb2010,Stark2014,Berg2016,Berg2019,
Vanzella2016,Vanzella2017,Senchyna2017,Senchyna2019b}, suggesting that the intense UV nebular emission can be explained as a natural byproduct of 
the hard radiation field associated with very young 
stellar populations that arise in metal poor galaxies with 
significant star formation. In the following section, we consider 
whether there are any sources in our sample which are unlikley 
to fit within this framework.


\begin{figure}
\begin{center}
\includegraphics[width=0.95\linewidth]{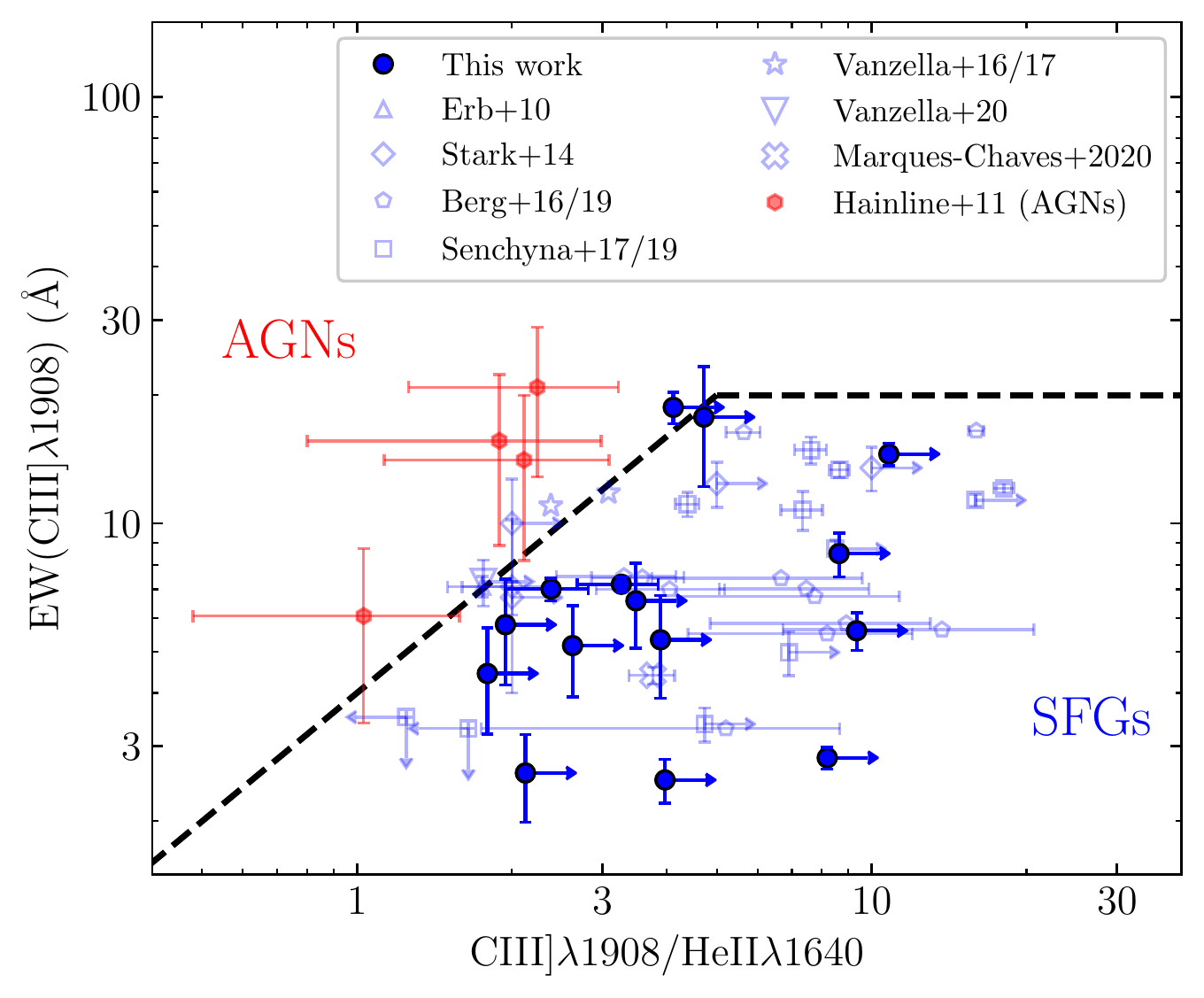}
\caption{C~{\scriptsize III}] EW versus C~{\scriptsize III}]/He~{\scriptsize II} diagnostic. The black dashed line shows the boundary between AGNs (upper left) and star-forming galaxies (SFGs; lower right). C~{\scriptsize III}] emitters in our sample are presented by blue circles. Blue open symbols show the data of metal-poor star-forming galaxies at $z=0-3$ from literature \citep{Erb2010,Stark2014,Berg2016,Berg2019,Senchyna2017,Senchyna2019b,Vanzella2016,Vanzella2017,Vanzella2020,Marques-Chaves2020}. Red hexagons show narrow-line AGNs from \citet{Hainline2011}.}
\label{fig:uv_ratio}
\end{center}
\end{figure}

\subsection{Ionizing Sources} \label{sec:ionizing_source}

In the last several years, work has focused on the 
development of rest-frame UV diagnostics that distinguish galaxies 
with spectra powered by stars from those powered by AGNs 
\citep[e.g.][]{Feltre2016,Gutkin2016,Nakajima2018}. Owing to the presence of several high ionization 
emission features (C~{\small IV}, He~{\small II}), the rest-frame UV is particularly 
well-suited to identify line ratios that require a power-law 
AGN spectrum in the extreme UV. Here we consider our emission 
line detections in the context of several of these diagnostics, 
with the goal of identifying any sources that 
are not satisfactorily explained with massive stars. This 
step is a necessary prerequisite for assessing the frequency of 
low metallicity stellar populations in galaxies with 
extreme [O~{\small III}] emission, while also providing some insight into 
how common AGN might be in similarly-selected samples at $z\gtrsim6$. 

In Figure \ref{fig:uv_ratio}, we compare our C~{\small III}] 
emitting galaxies to the C~{\small III}] EW 
versus C~{\small III}]/He~{\small II} diagnostic 
developed by \citet{Nakajima2018}. As can be seen in the figure, 
AGN are generally found with either larger C~{\small III}] EWs 
or smaller C~{\small III}]/He~{\small II} ratios, the 
latter reflecting the stronger He~{\small II} emission 
that arises from an AGN power-law spectrum. We use this diagnostic 
primarily because of the data we have available to us at present 
(our spectral coverage often does not extend to C~{\small IV}). 
Note that many of our C~{\small III}]/He~{\small II} 
measurements are lower limits, owing to non-detections of He~{\small II}.  
In addition to plotting the line 
ratios of our sample, we overlay those of AGNs \citep{Hainline2011} and 
metal poor star-forming galaxies in the literature \citep{Erb2010,Stark2014,Berg2016,Berg2019,Senchyna2017,Senchyna2019b,Vanzella2016,Vanzella2017}. The majority of galaxies in our 
sample lie in the part of the diagram associated with 
star-forming galaxies in the \citet{Nakajima2018} models. 
Comparison to the data in the literature 
gives a similar picture, with the bulk of the sample having either 
larger C~{\small III}]/He~{\small II} ratios ($\gtrsim2$) or lower 
C~{\small III}] EWs than the AGNs in \citet{Hainline2011}.  

However we note that the two strongest  C~{\small III}] emitters in our 
sample (EW$_{\rm{CIII]}}=18.7$~\AA\ for COSMOS-04432 and 
EW$_{\rm{CIII]}}=17.8$~\AA\ for UDS-07665) are situated very close to the AGN boundary of the \citet{Nakajima2018} diagnostic. For both sources, the [O~{\small III}] line is also extremely prominent, with 
EW$_{\rm{[OIII]+H\beta}}=1608$~\AA\ (COSMOS-04432) and 
EW$_{\rm{[OIII]+H\beta}}=1800$~\AA\ (UDS-07665).
These objects thus have nebular line properties that are very similar 
to what has been seen at $z>6$, but their position in Figure \ref{fig:uv_ratio} suggests 
that AGN ionization might be required to explain the line ratios. 
The blended C~{\small III}] doublets 
of both sources are narrower ($\Delta v=253\pm20$ km/s) than the 
C~{\small III}] emission seen in type I AGNs 
($\Delta v$ up to $\sim1000$ km/s; \citealt{LeFevre2019}) but similar 
to typical type II AGNs. While neither source has an X-ray counterpart within 
a $2.0$ arcsec radius, this is consistent with the low X-ray luminosities 
measured in the C~{\small III}]-emitting AGNs discussed in \citet{LeFevre2019}.

\begin{figure}
\begin{center}
\includegraphics[width=0.9\linewidth]{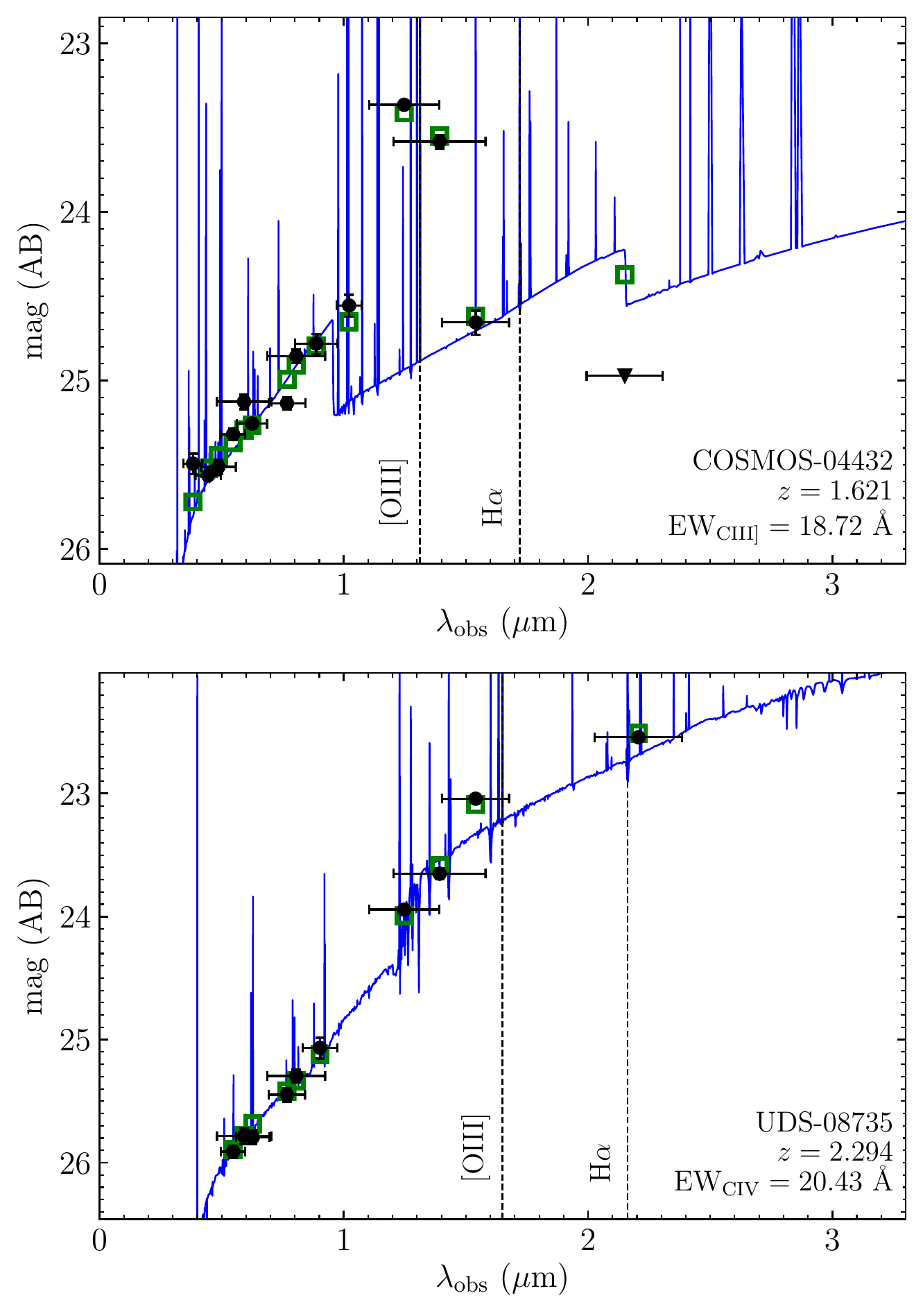}
\caption{Broadband SEDs of the two possible AGNs, the extreme C~{\scriptsize III}] emitter COSMOS-04432 (upper panel) and the extreme C~{\scriptsize IV} emitter UDS-08735 (lower panel), both with red UV slopes ($\beta>-1.0$). Observed broadband photometry is shown as solid black circles. The best-fit SED models inferred from BEAGLE are plotted by solid blue lines, and synthetic photometry is presented by open green squares.}
\label{fig:agn_sed}
\end{center}
\end{figure}

The broadband SEDs offer additional information. The data for UDS-07665 
are shown in the upper left panel of Figure \ref{fig:sed}. Its UV slope is blue ($\beta=-2.2$) and the near-infrared filters show the characteristic flux excess from strong [O~{\small III}]+H$\beta$ emission. Such SEDs are very common at 
$z>6$ and are consistent with expectations for an unreddened metal 
poor galaxy dominated by 
very young stellar populations. The SED of COSMOS-04432 (top panel of Figure \ref{fig:agn_sed}) 
appears very different. While the flux excess from the rest-frame optical nebular 
lines are present, the rest-frame UV continuum appears very red ($\beta=-0.90$), 
likely implying a significant dust covering fraction.  Such red UV slopes 
are very similar to what is often seen in UV-selected $z\simeq3$ galaxies with 
narrow-lined AGN \citep{Hainline2011,LeFevre2019}. While rare 
at $z>6$, red UV colors are present in a subset of the population 
with intense [O~{\small III}] emission \citep{Smit2018}. COSMOS-04432 appears 
to be an analog of these reddened reionization-era sources. Deeper 
spectra of this low-z analog should reveal detection of the higher 
ionization lines (C~{\small IV}, He~{\small II}) necessary to clarify whether a 
power-law AGN spectrum is indeed present.

The flux ratio of C~{\small IV} and O~{\small III}] provides 
an additional way to identify sources that may be powered by 
AGN (e.g., \citealt{Feltre2016,Mainali2017}). In particular, 
the intense radiation field powered by an AGN spectrum triply ionizes 
a significant fraction of the oxygen, leading to weaker O~{\small III}] 
and smaller O~{\small III}]/C~{\small IV} flux ratios. Our sample contains five objects with nebular C~{\small IV} measurements
(COSMOS-04064, COSMOS-22402, UDS-07665, UDS-08735, UDS-19167), none of 
which show X-ray detections at their positions.  
Four of the five systems (all but UDS-08735) exhibit line ratios 
in the range $\log($O~{\small III}]/C~{\small IV}$)=-0.12$ to $0.13$, consistent with observations of metal poor star-forming 
galaxies \citep[e.g.][]{Stark2014,Berg2016,Vanzella2016}. The BEAGLE models suggest a similar picture for these four galaxies, 
with low metallicities ($Z=0.04-0.28\ Z_\odot$), young stellar 
populations ($4-26$~Myr for constant star formation) and large ionization 
parameters ($\log{U}=-1.4$ to $-2.2$) required to reproduce the 
full spectra and photometry.  

The rest-frame UV spectrum of UDS-08735 is more challenging to reproduce with 
stars as the primary ionizing source. The C~{\small IV} EW is the largest in 
our sample ($20.4\pm4.7$~\AA), one of the only measurements known 
with similar values to those seen at $z>6$. As with the sources 
described above, no X-ray source is present in the vicinity 
of the galaxy.  However the $\log($O~{\small III}]/C~{\small IV}$)$ ratio ($<-0.67$ at $3\sigma$) is much lower than in the systems described above, similar instead to the values 
observed in $z\sim2-4$ type II quasars ($<-0.4$, \citealt{Hainline2011, Alexandroff2013}). The SED of UDS-08735 (bottom panel of Figure \ref{fig:agn_sed}) reveals several additional differences. The UV slope is
very red ($\beta=-0.46$), which as we described above 
is comparable to most AGNs which contaminate star-forming galaxy samples \citep{Hainline2011,LeFevre2019}. 
The flux excesses from the rest-frame optical 
lines are much weaker than in most of our galaxies, reflecting one of the 
lower [O~{\small III}]+H$\beta$ equivalent widths in our sample (EW = $460\pm24$~\AA). 
If powered by stars, this would indicate a much older stellar population 
(and hence softer radiation field) than is often linked to strong C~{\small IV} (e.g., \citealt{Senchyna2019b}). Indeed, the tension in reproducing  UDS-08735
with stellar photoionization is readily apparent with BEAGLE. The 
best-fit C~{\small IV} EW  ($0.3^{+0.4}_{-0.2}$~\AA) is nearly 70$\times$ lower 
than what is observed. We conclude that this source is most likely an 
AGN and do not include it in the analysis in the following section.  


\begin{figure}
\begin{center}
\includegraphics[width=\linewidth]{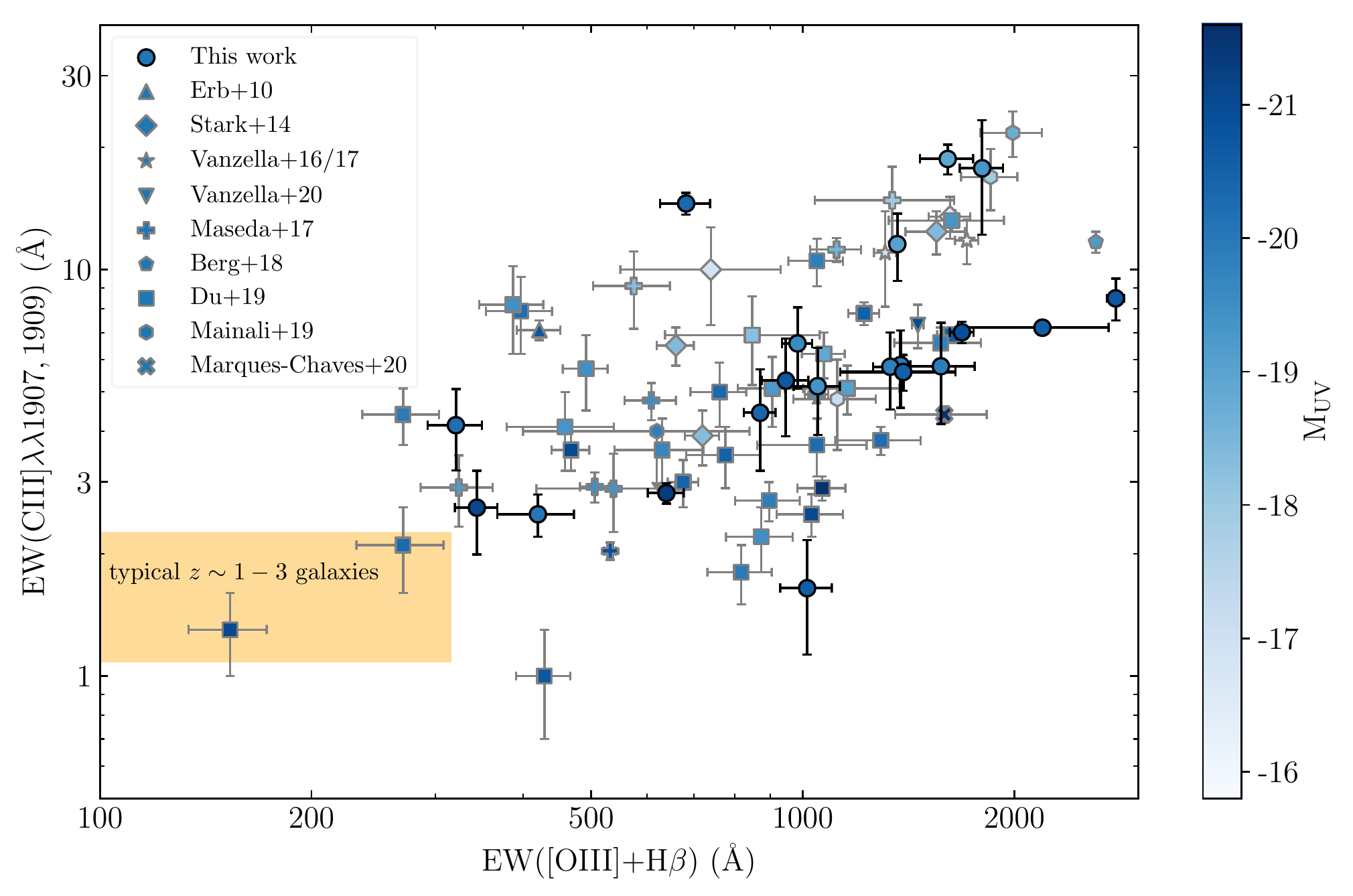}
\caption{C~{\scriptsize III}] EW as a function of [O~{\scriptsize III}]+H$\beta$ EW for extreme optical line emitters at $z\sim2-3$, with color indicating the absolute UV magnitude. Data from objects in our spectroscopic sample are shown as circles. Data from previous investigations of C~{\scriptsize III}] emission in $z\sim2-3$ galaxies are also included \citep{Erb2010,Stark2014,Vanzella2016,Vanzella2017,Maseda2017,Berg2018,Vanzella2020,Mainali2020,Du2020,Marques-Chaves2020}. The orange shaded region represents the average and $68\%$ confidence interval of C~{\scriptsize III}] EW \citep{Shapley2003} and [O~{\scriptsize III}]+H$\beta$ EW \citep{Reddy2018} of typical, more massive $z\sim1-3$ galaxies.}
\label{fig:c3ew_o3hbew}
\end{center}
\end{figure}


\begin{figure*}
\begin{center}
\includegraphics[width=0.9\linewidth]{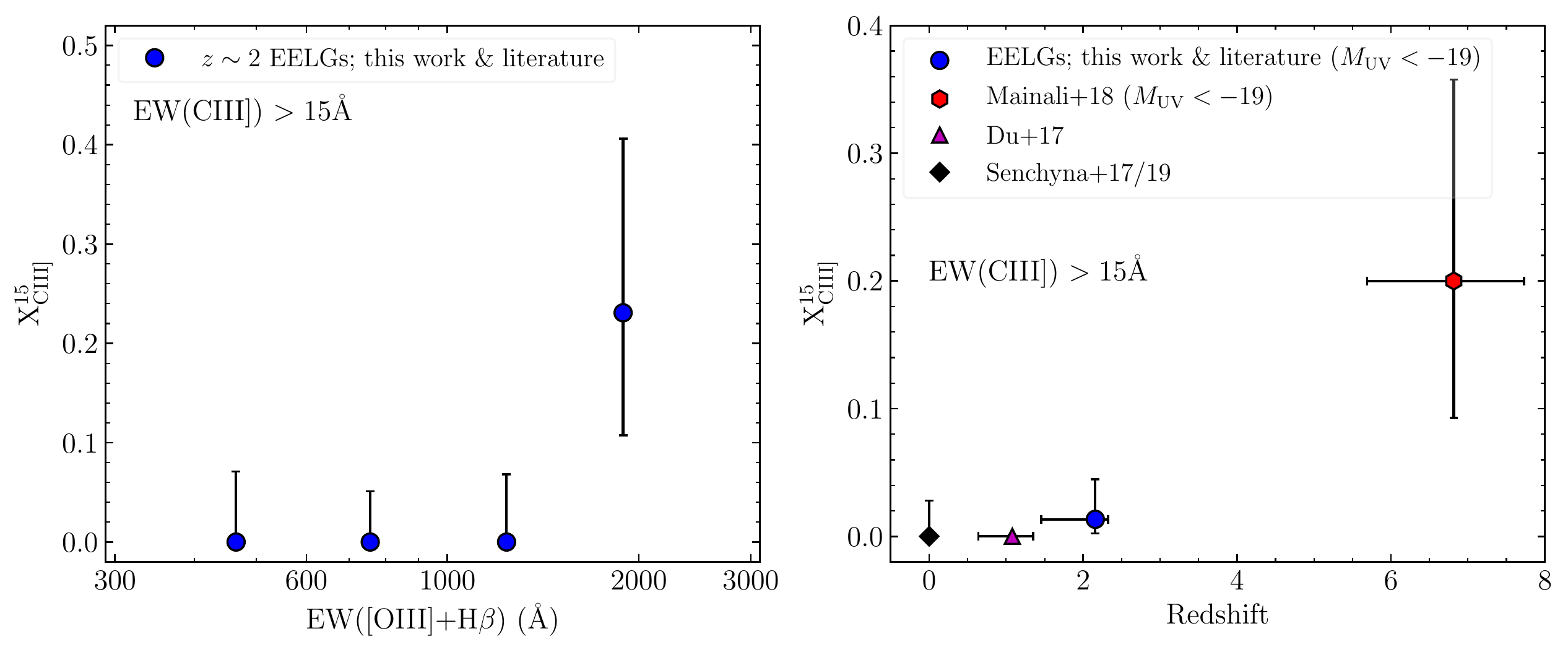}
\caption{Left panel: Fraction of galaxies with extremely large C~{\scriptsize III}] EWs ($>15$~\AA) as a function of [O~{\scriptsize III}]+H$\beta$ EW. We show the C~{\scriptsize III}] fractions for four [O~{\scriptsize III}]+H$\beta$ EW bins ($=300-600$~\AA, $600-1000$~\AA, $1000-1500$~\AA, and $1500-3000$~\AA) for a sample combining our $z=1.3-3.7$ EELGs and the $z\sim2$ EELGs in \citet{Du2020} and \citet{Mainali2020}. Right panel: Fraction of galaxies with C~{\scriptsize III}] EW $>15$~\AA\ as a function of redshift. The fraction at $z\sim0$ (black diamond) and $z\sim1$ (magenta triangle) are derived from \citet{Senchyna2017,Senchyna2019b} and \citet{Du2017}. The $z>5.4$ data (red hexagon) is computed using the spectroscopic sample from literature, which are summarized in \citet{Mainali2018} and add the newly detected C~{\scriptsize III}] in a $z=7.5$ galaxy \citep{Hutchison2019}. The fractions of extreme C~{\scriptsize III}] emitter at $z\sim2$ in EELGs (combining our sample and those in \citealt{Du2020} and \citealt{Mainali2020}; blue circle) are constrained to objects with [O~{\scriptsize III}]+H$\beta$ EW $=300-3000$~\AA\ and $M_{\rm{UV}}<-19$, which are comparable to the values of extreme C~{\scriptsize III}] emitters at $z>5.4$.} 
\label{fig:c3_frac}
\end{center}
\end{figure*}


\begin{figure*}
\begin{center}
\includegraphics[width=0.9\linewidth]{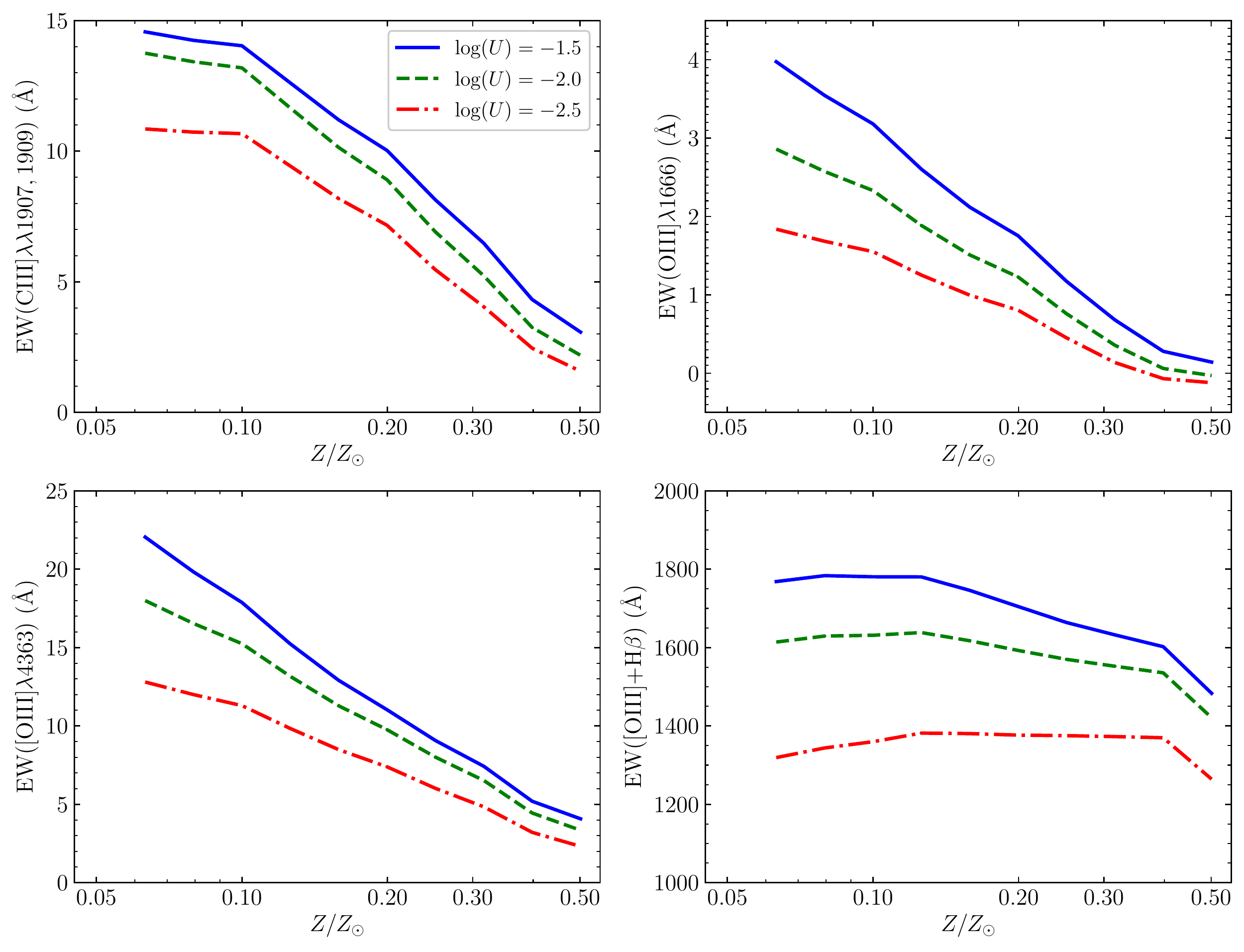}
\caption{Predicted C~{\scriptsize III}]~$\lambda\lambda1907,1909$ EW (upper left), O~{\scriptsize III}]~$\lambda1666$ EW (upper right), [O~{\scriptsize III}]~$\lambda4363$ EW (lower left), and [O~{\scriptsize III}]+H$\beta$ EW (lower right) from BEAGLE models for different metallicities at age $=10$~Myr (assuming constant star formation history). Models with different ionization parameters are shown by red dash-dotted lines ($\log{U}=-2.5$), green dashed lines ($\log{U}=-2.0$), and blue solid lines ($\log{U}=-1.5$).}
\label{fig:model_Z}
\end{center}
\end{figure*}

\section{Discussion} \label{sec:discussion}

Over the last decade, spectroscopic searches have begun to 
discover a handful of reionization-era analogs with rest-frame UV 
properties similar to those seen at $z>6$. In this paper, 
we have taken steps toward assembling a statistical sample 
of such systems at $z\simeq2-3$. Here we use this database 
for two purposes. First, we investigate the C~{\small III}] EW distribution 
in galaxies with [O~{\small III}]+H$\beta$ EW $=1000-2000$~\AA, with 
the goal of testing whether the C~{\small III}] 
properties seen in the reionization era (C~{\small III}] EW $>15$~\AA) 
are uniformly present in low mass galaxies dominated by extremely 
young stellar populations (Section \ref{sec:c3_strength}). Second, we quantify the fraction of 
star-forming galaxies in our sample with strong UV line emission,  
with the goal of constraining how common 
low metallicity stellar populations are in low mass 
galaxies with large sSFRs at $z\simeq2-3$ (Section \ref{sec:redshift_evolution}). Both measurements 
will provide a $z\simeq2-3$ baseline against which current and 
future $z>6$ observations can be compared.  

\subsection{A Baseline for interpreting $z>7$ C~{\small III}] detections} \label{sec:c3_strength}

Our first goal is geared toward understanding the high EW 
($>15$~\AA) C~{\small III}] detections at $z>6$. While AGNs 
have been discovered with similar UV line properties at 
lower redshifts \citep{LeFevre2019}, there are very few examples of 
star-forming galaxies powering comparable spectra, limiting our 
ability to interpret the nebular emission that has been detected 
at $z>6$. \citet{Mainali2020}  
recently presented discovery of two $z\simeq2$ gravitationally 
lensed galaxies with C~{\small III}] EW $\simeq15-20$~\AA, each of which was selected 
to have the intense [O~{\small III}]+H$\beta$ emission ($\simeq1500-2000$~\AA)  
that is characteristic of galaxies dominated by very young stellar 
populations. Such systems are extremely rare at lower redshifts, 
so little is known about the range of rest-frame UV line spectra they exhibit. 

Following the survey presented in this paper 
as well as those in the literature that also satisfy our selection criteria \citep{Mainali2020,Du2020}, 
there is now rest-frame UV spectral coverage of C~{\small III}] for $69$ systems with 
[O~{\small III}]+H$\beta$ EW $>1000$~\AA, including $12$ 
systems with [O~{\small III}]+H$\beta$ EW $>2000$~\AA.
In Figure \ref{fig:c3ew_o3hbew}, we plot the dependence of C~{\small III}] EW on the 
[O~{\small III}]+H$\beta$ EW for galaxies in our sample and those in the literature.  
As has been seen previously  
(e.g., \citealt{Senchyna2017,Senchyna2019b,Maseda2017,Du2020}), the largest EW C~{\small III}] 
emission is seen predominantly at the largest rest-frame optical line EWs.  
The data in Figure \ref{fig:c3ew_o3hbew} show that C~{\small III}] detections with 
EW $>10$~\AA\ only begin to be seen in galaxies with 
[O~{\small III}]+H$\beta$ EW $>500$~\AA.  While such intense optical 
line emission is relatively rare among massive star 
forming galaxies at $z\simeq2-3$ \citep[e.g.][]{Reddy2018}, 
it becomes the norm among galaxies in the reionization era 
(e.g., \citealt{Labbe2013,Smit2014,Smit2015,deBarros2019,Endsley2020}).  

We also see in Figure \ref{fig:c3ew_o3hbew} that the C~{\small III}] EWs that have been detected 
in $z>6$ galaxies ($>15-20$~\AA) appear only at [O~{\small III}]+H$\beta$ EW $>1500$~\AA\ 
in our $z\simeq 2-3$ sample.  
Given the large sSFR required to power these optical line 
EWs \citepalias{Tang2019}, this suggests that the C~{\small III}] lines may be a 
natural consequence of the hard radiation field associated with extremely young stellar 
populations ($<5-10$~Myr). This 
trend is most clearly seen in the left panel of Figure \ref{fig:c3_frac}, where we 
plot the fraction of galaxies observed in our sample and the literature with C~{\small III}] 
EW $>15$~\AA\ as a function of [O~{\small III}]+H$\beta$ EW. The C~{\small III}] 
emitter fraction is defined in our analysis as the ratio of sources with C~{\small III}] EW $>15$~\AA\ 
and the total number of galaxies for which we obtained C~{\small III}] measurements. 
The latter quantity includes only those non-detections for which we can rule out 
C~{\small III}] with EW $>15$~\AA\ (i.e., we do not include spectra 
that are too shallow to detect strong emission).  
Of the $86$ galaxies we have observed with 
sufficient sensitivity at [O~{\small III}]+H$\beta$ EW $=300-1500$~\AA, 
none have been found to exhibit C~{\small III}] 
above this threshold. In the $10$ galaxies we have targeted with 
[O~{\small III}]+H$\beta$ EW $>1500$~\AA, we find $1$ galaxy 
that has C~{\small III}] EW $>15$~\AA\ 
(here we have excluded COSMOS-04432, the candidate AGN 
discussed in Section \ref{sec:ionizing_source}), 
implying a fraction of $0.10^{+0.19}_{-0.08}$. 
Error bars are derived by adopting the statistics from \citet{Gehrels1986}. 
If we include sources in \citet{Du2020} and \citet{Mainali2020}, 
this fraction increases to $0.23^{+0.18}_{-0.12}$. These results suggest that 
the C~{\small III}] EWs seen in the 
reionization era ($>15$~\AA) require very strong optical line emission, 
becoming somewhat more common in galaxies with 
[O~{\small III}]+H$\beta$ EW $>1500$~\AA, similar to  
previous results at $z\simeq0$ and $z\simeq2$ \citep{Senchyna2017,Mainali2020,Du2020}. 

What is perhaps most striking about the results described above is 
that intense C~{\small III}] emission is 
not found to be the norm among the most extreme optical line emitters.
Indeed, four of the five galaxies in our [O~{\small III}]+H$\beta$ EW $>1500$~\AA\ 
sample have C~{\small III}] EW below $10$~\AA.  
This includes the most intense line emitter UDS-29267 
([O~{\small III}]+H$\beta$ EW $=2788\pm80$~\AA) in our sample.  
Based on its optical lines, we naively expected to see among the strongest 
C~{\small III}] in our sample. However the spectrum revealed a 
detection with C~{\small III}] EW $=8.5$~\AA, well below the 
values that have been seen at $z>6$.  Clearly the hard radiation field that 
emerges from young stellar populations is not a sufficient condition for 
intense C~{\small III}] emission. 

The BEAGLE photoionization models suggest that the spread in 
C~{\small III}] at [O~{\small III}]+H$\beta$ EW $>1500$~\AA\ 
is driven by metallicity. As detailed in Section \ref{sec:modeling}, the modeling 
results are based on fits to the rest-frame UV and optical nebular lines 
together with the broadband continuum, and here we focus on those sources 
with extreme [O~{\small III}]+H$\beta$ EW ($>$1500~\AA).  The results indicate
that those galaxies in this subset with strong C~{\small III}] ($11.6-17.8$~\AA) 
tend to require lower metallicities (median of $0.07\ Z_\odot$), whereas those 
with weaker C~{\small III}] ($5.8-7.2$~\AA) are fitted with more moderate metallicities (median of $0.28\ Z_\odot$). The strong and weak C~{\small III}] emitters are found to have nearly-identical ionization parameters (median of $\log{U}=-1.71$ and $\log{U}=-1.68$, respectively) and fairly similar specific star formation 
rates (median of $79$~Gyr$^{-1}$ and $129$~Gyr$^{-1}$). Both subsets 
prefer sub-solar C/O ratios ($0.52$ C/O$_\odot$), and deviations from this 
lead to tension with the observed C~{\small III}] strength.   
The photoionization models thus suggest that metallicity is the primary 
driver of the spread in C~{\small III}] at fixed [O~{\small III}] EW.  
More work is needed to confirm this trend with the temperature-sensitive auroral lines. For the few existing cases in which 
high S/N auroral line detections exist, the direct-method metallicities 
confirm the presence of low metallicities ($\simeq10\%\ Z_\odot$) in the 
most extreme C~{\small III}] emitters (see e.g., \citealt{Mainali2020}).  

The luminosity distribution of our galaxies suggests a similar picture, 
with the strength of C~{\small III}] emission (at fixed [O~{\small III}]+H$\beta$ EW) 
generally increasing toward lower luminosities (see color bar in Figure \ref{fig:c3ew_o3hbew}) 
where low metallicity gas and stars are more likely. 
To investigate the dependence of C~{\small III}] on UV luminosity, we 
quantify the average C~{\small III}] EW in bins of $M_{\rm{UV}}$ for 
galaxies with extreme [O~{\small III}]+H$\beta$ emission 
(EW $>1500$~\AA), including those in our sample and those in the 
literature selected similarly \citep{Mainali2020,Du2020}. To avoid biasing our results, we only 
consider those spectra with the sensitivity to detect 
weak C~{\small III}] emission (EW $=5$~\AA).  As we will discuss 
below, this limits our sample of lower luminosity galaxies where 
continuum magnitudes are faint. Focusing first on luminous galaxies ($M_{\rm{UV}}<-19.5$), 
there are four [O~{\small III}]+H$\beta$ 
EW $>1500$~\AA\ systems in our sample with deep C~{\small III}] 
constraints.  Of these, none have C~{\small III}] EW $>10$~\AA, and 
the median value is just C~{\small III}] EW $=6.5$~\AA.  
At lower luminosities ($M_{\rm{UV}}>-19.5$), there are only three 
[O~{\small III}]+H$\beta$ EW $>1500$~\AA\ galaxies with 
deep enough spectra to detect weak C~{\small III}], allowing a 
fair comparison with our bright sample. Two of these are from \citet{Mainali2020} and the third is from \citet{Du2020}. Notably, 
all three have C~{\small III}] EW $>10$~\AA, with a median 
value of C~{\small III}]  EW $=16.9$~\AA.  While samples 
are admittedly small, these results suggest that  at $z\simeq2-3$, 
the strongest C~{\small III}] emission (EW $>10$~\AA) is  
found primarily in low luminosity galaxies.  

The role of metallicity in regulating C~{\small III}] has been 
pointed out elsewhere \citep{Erb2010,Stark2014,Rigby2015,Jaskot2016,Senchyna2017,Byler2018,Plat2019,Du2020,Ravindranath2020}, reflecting increased collisional excitation in higher 
temperature gas and the harder radiation field associated with metal poor massive stars. But what is perhaps surprising about our results is that C~{\small III}] appears to react more 
to a change in metallicity than [O~{\small III}] over the range considered here ($0.1$ to $0.3\ Z_\odot$).  This can be seen more clearly in Figure \ref{fig:model_Z}, where we plot the dependence of the 
C~{\small III}] EW on metallicity in BEAGLE at fixed stellar age ($10$~Myr, assuming constant star formation history) and fixed ionization parameter.  
Since ionization parameter is well-known to anti-correlate with metallicity, 
we show the trend for three ionization parameters ($\log{U}=-2.5$, $-2.0$, $-1.5$).  
We also display the same trends for two temperature-sensitive auroral lines 
(O~{\small III}]~$\lambda1666$ and [O~{\small III}]~$\lambda4363$) and 
for the [O~{\small III}]+H$\beta$ optical lines.  Between $0.3$ and $0.1\ Z_\odot$, we 
see that the C~{\small III}] EW increases by roughly a factor of two 
(at fixed $\log{U}$).  This closely tracks the rise seen in both auroral lines, 
suggesting that C~{\small III}] is very sensitive to the gas temperature 
in this metallicity regime. 
In contrast, the [O~{\small III}]+H$\beta$ EW remains nearly constant over the metallicity range.  Even accounting for the 
larger ionization parameters expected at lower metallicity ($\log{U}=-2.0$ at $0.1\ Z_\odot$ and $\log{U}=-2.5$ at $0.3\ Z_\odot$ according to 
\citealt{Perez-Montero2014} and \citealt{Sanders2020}), the BEAGLE 
models predict that the [O~{\small III}]+H$\beta$ EW will only increase by a 
factor of $1.1\times$ over this metallicity range. While this is computed for stellar ages of $10$~Myr, we find similar results for older age populations.  A similar 
picture was previously discussed in \citet{Jaskot2016} (see their Figure 9). 
Because we are plotting equivalent width, the trend seen in Figure \ref{fig:model_Z} is sensitive to 
how the underlying continuum varies with metallicity.  Since bluer colors are produced by 
lower metallicity stellar populations, we expect a slight decline in 
optical EWs with respect to those seen in the UV as we look toward 
lower metallicities.  However, we have verified using BEAGLE that the 
trends seen in Figure \ref{fig:model_Z} remain when considering only 
line luminosity (see also \citealt{Jaskot2016}), suggesting that 
continuum is not a dominant factor. 

The physical explanation for the divergent metallicity trends of the 
C~{\small III}] and [O~{\small III}] luminosities is 
straightforward and has been suggested previously in \citet{Plat2019}. 
The emissivity of both lines is proportional to the 
collisional excitation 
rate, which in turn depends on the ionized gas temperature and the energy 
of the excited state as $T^{-1/2} e^{-E/kT}$.  For C~{\small III}]~$\lambda\lambda1907,1909$, 
the excited state has an energy of $E/k=7.6\times10^4$ K, while for 
[O~{\small III}]~$\lambda\lambda4959,5007$, the energy is just 
$E/k=2.9\times10^4$ K.  It is straightforward to see that as the gas temperature 
increases from $1.0$ to $1.5\times10^4$ K, the emissivity of C~{\small III}] increases considerably 
more than that of [O~{\small III}].
The two auroral lines shown in Figure \ref{fig:model_Z} have excited 
states with energies similar to that of C~{\small III}] 
($E/k=8.7\times10^4$ K for O~{\small III}]~$\lambda1666$ and $E/k=6.2\times10^4$ K for [O~{\small III}]~$\lambda4363$), leading to a similar trend with metallicity.  These 
results suggest that as we approach the metallicity range of $0.1-0.3\ Z_\odot$ 
(an important regime at $z>6$; e.g., \citealt{Jones2020}), 
the collisionally-excited UV lines will play an increasingly important 
role in cooling owing to the higher energies of their excited states \citep{Plat2019} and as a result, should be increasingly prominent in deep rest-frame UV spectra. It is conceivable that 
early galaxies will be significantly enhanced in oxygen relative to iron 
\citep[e.g.][]{Steidel2016,Strom2018,Sanders2020,Jeong2020,Topping2020}, leading to a harder ionizing spectrum and potentially 
stronger nebular emission. Following the same photoionization modeling approach as \citet{Topping2020}, we 
find that variations in stellar metallicity at a given gas-phase metallicity 
do not significantly change the strength of C~{\small III}] at fixed 
[O~{\small III}] EW. The dispersion in C~{\small III}] at a given [O~{\small III}] 
EW appears to be dominated by the variations in gas-phase metallicity described above. 

The results laid out above have implications for our understanding 
of the C~{\small III}] detections at $z>7$ \citep{Stark2017,Hutchison2019}. 
In addition to having extremely strong C~{\small III}] (EW $>16-22$~\AA), 
these galaxies are found to be quite luminous ($M_{\rm{UV}}=-22.1$ and $-21.6$).
As we outlined above, at $z\simeq2-3$, very strong C~{\small III}] is primarily 
seen among lower luminosity ($M_{\rm{UV}}>-19.5$) galaxies, as it is in these 
systems where the metal poor gas required to support such intense emission
is more likely to be found. The presence of strong C~{\small III}] 
emission in luminous galaxies at $z\simeq7-8$ may be one of the first hints 
of evolution in the luminosity-metallicity relationship between $z\simeq2-3$ and 
$z\simeq7$, with lower metallicities at fixed $M_{\rm{UV}}$ in the 
reionization-era population. Current sensitivity limits at $z\simeq7$ 
allow us to detect C~{\small III}] only in galaxies with extreme [O~{\small III}]+H$\beta$ 
emission, so these inferences are valid only for the subset of the population 
with very large sSFR. The emergence of new samples of bright galaxies at $z\simeq7$ with intense optical line emission (e.g., \citealt{Endsley2020}) should allow 
improved statistics on the  C~{\small III}] EW distribution in this extreme population in the near future.  

\subsection{Redshift evolution of UV line fractions} \label{sec:redshift_evolution}

The results presented in this paper allow us to revisit the redshift 
evolution of the fraction of galaxies with intense (EW $>15$~\AA) 
C~{\small III}] emission (right panel of Figure \ref{fig:c3_frac}). 
Here we have updated the $z>5.4$ 
datapoint from \citet{Mainali2018} to include the new detection 
presented in \citet{Hutchison2019} and reconsider the measurement 
at $z\simeq2-3$ (see below). While statistics are 
still limited, it is clear that the fraction of galaxies with 
strong C~{\small III}] (EW $>15$~\AA) at $z>5.4$ ($0.20^{+0.16}_{-0.11}$) is 
larger than in samples at lower redshift. Some of this evolution can be explained by recent results which demonstrate the 
emergence of a significant population of extreme sSFR galaxies 
($>200$~Gyr$^{-1}$) with [O~{\small III}]+H$\beta$ EW $>1500$~\AA\ at $z>6$ \citep{deBarros2019,Endsley2020}, as it is 
only above these large optical line EWs where intense C~{\small III}] emission 
becomes common in star-forming galaxies (e.g., left panel of Figure 9). 

However if the evolution toward larger [O~{\small III}]+H$\beta$ EWs 
is the entire explanation for the rise in the frequency 
of detecting intense C~{\small III}] emission, then we would expect the C~{\small III}] fraction in our sample of $z\simeq 2-3$ EELGs to be similar to that seen 
in the reionization era. To test this, we have attempted to construct a sample 
that is matched in [O~{\small III}]+H$\beta$ EW to that 
in the $z>5.4$ sample. We include galaxies with 
[O~{\small III}]+H$\beta$ EW $=300-3000$~\AA\ and $M_{\rm{UV}}<-19$, including our 
sample ($46$ sources) together the [O~{\small III}]-selected samples that 
satisfy these criteria in \citet{Du2020} ($28$ sources) and 
\citet{Mainali2020} ($1$ source). The median [O~{\small III}]+H$\beta$ 
EW in this sample is $760$~\AA, similar 
to that in samples at $z\simeq7-8$ \citep{Labbe2013,deBarros2019,Endsley2020}.  
Of the $75$ galaxies that satisfy these criteria, only $1$ has C~{\small III}] EW $>15$~\AA. 
This implies a fraction of $0.013^{+0.031}_{-0.011}$, much lower than seen at $z>5.4$. 
While the uncertainties at $z>5.4$ remain sizeable, this suggests that the emergence 
of extreme sSFR systems at $z\simeq7$ \citep{Endsley2020} is not enough 
on its own to drive the large C~{\small III}] emission we have observed in early galaxies. As described in Section \ref{sec:c3_strength}, we suggest that this may additionally reflect evolution toward lower metallicities at earlier times. Indeed,  
if the luminous galaxy population evolves from $0.5$ to $0.1\ Z_\odot$ between $z\simeq 2$ and $z\simeq 7$, 
we would expect to see a significant rise (up to $5\times$) in C~{\small III}] EWs without any evolution in sSFR (Figure \ref{fig:model_Z}). We thus suggest that the combination of larger sSFRs and lower metallicities likely work together to explain the large C~{\small III}] EWs that are being observed at $z>6$.

Finally we compare the fraction of star-forming sources in our 
$z\simeq2-3$ sample with nebular C~{\small IV} emission to that seen at $z>6$.  
Since powering C~{\small IV} requires a hard ionizing spectrum associated 
with very low metallicity stars ($<0.1\ Z_\odot$), this 
comparison will allow us to investigate evolution in the 
incidence of very low metallicity stellar populations among 
galaxies with extreme optical line emission. The first 
step is to assemble the statistical baseline at $z\simeq2-3$. 
Here we consider objects with large [O~{\small III}]+H$\beta$ 
EWs ($=1000-3000$~\AA) and $M_{\rm{UV}}<-19$, focusing on 
galaxies at $z=1.7-3$, where C~{\small IV} is visible from the 
optical spectrographs we have used. Taking together our sample and 
the extreme [O~{\small III}] emitter sample in \citet{Du2020} 
(and removing the AGN candidate discussed in Section \ref{sec:ionizing_source}), we find that only $2$ 
out of $21$ sources at $z\sim2-3$ show C~{\small IV} EW $>5$~\AA, 
and none of them shows C~{\small IV} EW $>10$~\AA. For very low metallicity systems,
we might begin to see weaker [O~{\small III}] emission, so we also consider 
galaxies with [O~{\small III}]+H$\beta$ EW $=500-1000$~\AA. Here there are another 
$23$ objects from both our sample and that of \citet{Du2020} that have 
deep C~{\small IV} constraints, but none show detections of C~{\small IV} with EW $>5$~\AA.

These results stand in sharp contrast to the two C~{\small IV} detections 
at $z>6$, both of which show EW $>20$~\AA\ in $z>6$ galaxies with 
$M_{\rm{UV}}=-20$ to $-19$ \citep{Stark2015b,Mainali2017,Schmidt2017}. The 
absence of such line emission in similar luminosity systems at $z\simeq2-3$ 
is suggestive of a shift toward some combination of harder ionizing spectra 
and weaker underlying UV continuum in a subset of $z>6$ galaxies.  
Since we are considering samples of similar optical line EW (and similar sSFR), this cannot simply 
be explained by evolution towards galaxies dominated by younger 
stellar population (i.e., larger [O~{\small III}]+H$\beta$ EW). 
Instead, this likely implies the emergence of massive-star populations dominated 
by very low metallicities at early times. Ultimately larger datasets at $z>6$ will 
present a  clearer picture of the C~{\small IV} EW 
distribution in the reionization era and the physical cause of any 
evolution that is present. Statistical samples at $z\simeq2-3$ 
will continue to be critical in the interpretation of this population.


\section{Summary} \label{sec:summary}

We have presented rest-frame UV spectroscopic observations of $138$ galaxies with extreme [O~{\small III}]+H$\beta$ line emission at $z=1.3-3.7$ and $11$ sources at $z\sim4-6$. The galaxies have rest-frame [O~{\small III}]+H$\beta$ EWs $=300-3000$~\AA, similar to the EWs of galaxies in the reionization era \citep[e.g.][]{Labbe2013,Smit2015,Endsley2020}. The dataset allows us to measure the C~{\small III}]~$\lambda\lambda1907,1909$ and C~{\small IV}]~$\lambda\lambda1548,1550$ emission lines, which are found to be intense (EW $>15-20$~\AA) in $z>6$ galaxies \citep{Stark2015a,Stark2015b,Stark2017,Mainali2017,Hutchison2019}. Using this statistical sample of reionization-era analogs at $z\simeq1-3$, we aimed to interpret the nebular line emission emerging at $z>6$ with two goals. First, we investigated the C~{\small III}] EW distribution in galaxies with [O~{\small III}]+H$\beta$ EW $>1000-2000$~\AA, to see whether the EW $>15$~\AA\ C~{\small III}] line emission are uniformly present in these systems. Second, we sought to constrain the fraction of strong UV line emitters in our sample of large sSFR galaxies, providing a baseline at $z\simeq1-3$ to compare with future observations at $z>6$. We summarize our findings below:

(1) We detect rest-frame UV emission lines (C~{\small IV}~$\lambda\lambda1548,1550$, He~{\small II}~$\lambda1640$, O~{\small III}]~$\lambda\lambda1661,1666$, Si~{\small III}]~$\lambda\lambda1883,1892$, C~{\small III}]~$\lambda\lambda1907,1909$, or Mg~{\small II}~$\lambda\lambda2796,2803$) in $24$ extreme [O~{\small III}] emitters at $z=1.3-3.7$ in our sample using Magellan/IMACS and MMT/Binospec. Four objects show C~{\small IV} emission lines (EW $=1.5-20.4$~\AA) with S/N $>3$, with one of 
the C~{\small IV} detections coming from \citet{Du2020}. The blended C~{\small III}] doublet has been detected in $20$ sources, with EW $=1.6-18.7$~\AA\ and a median value of $5.8$~\AA. The C~{\small III}] EWs measured in our sample are consistent with those of $z\sim2$ EELGs in the literature \citep{Stark2014,Maseda2017,Du2020}, and are $\sim4\times$ larger than those of more massive, typical star-forming galaxies at $z\sim1-3$ \citep[e.g.][]{Shapley2003,Du2017}. The largest C~{\small IV} and C~{\small III}] EWs ($\simeq20$~\AA) found in our $z=1.3-3.7$ sample are comparable to that found at $z>6$.

(2) We find that for all but one of the $24$ galaxies with UV metal line detections at $z=1.3-3.7$, the nebular emission lines and broadband photometry can be reproduced by photoionization models powered by stars. The models suggest these systems are characterized by low stellar masses (median M$_{\star}=4.4\times10^8$ M$_{\odot}$), little dust 
reddening (median $E(B-V)=0.12$), large ionization parameters (median $\log{U}=-1.7$), and low metallicities (median $Z=0.2\ Z_\odot$). Both the observed C~{\small III}]/O~{\small III}] ratios and photoionization models imply sub-solar carbon-to-oxygen ratios (C/O $\simeq0.2-0.5$ C/O$_{\odot}$). The large [O~{\small III}]+H$\beta$ EWs (median value of $1080$~\AA) indicate very large sSFRs (median $=46$~Gyr$^{-1}$), suggesting an ionizing spectrum 
dominated by very young stellar populations. 

(3) We explore the relationship between C~{\small III}] EW and [O~{\small III}]+H$\beta$ EW for galaxies at $z\simeq1-3$. The intense C~{\small III}] line emission detected at $z>6$ (EW $>15$~\AA) only appears in galaxies with the largest [O~{\small III}]+H$\beta$ EWs ($>1500$~\AA), suggesting that very large sSFR ($\gtrsim100$~Gyr$^{-1}$)
is required to power the line emission seen at $z>6$. We find that there is significant 
variation in C~{\small III}] EW at fixed [O~{\small III}]+H$\beta$ EW, with many of the most extreme optical line 
emitters showing relatively weak C~{\small III}] emission (EW $<8$~\AA).  

(4) Photoionization models suggest that the spread 
in C~{\small III}] is driven by metallicity variations 
at fixed [O~{\small III}]+H$\beta$ EW, a result of the 
extreme sensitivity of C~{\small III}] to electron temperature.  We find that strong 
C~{\small III}] emission (EW $>15$~\AA) tends to be 
found in low metallicity galaxies ($\simeq0.1\ Z_\odot$) with 
large sSFR, and weaker C~{\small III}] (EW $=5-8$~\AA) tends 
to be found in moderate metallicity systems ($\simeq0.3\ Z_\odot$) 
with large sSFR.  The luminosity distribution supports this picture, with strong C~{\small III}] emission generally 
associated with the lowest luminosity ($M_{\rm{UV}}>-19.5$) 
galaxies at a given [O~{\small III}]+H$\beta$ EW, where low 
metallicities are most likely. Among more luminous 
galaxies at $z\simeq2-3$ (M$_{\rm{UV}}<-19.5$), we tend to find weaker C~{\small III}] emission.  These results 
stand in contrast to those at $z>7$, where very strong 
C~{\small III}] has already been identified in extremely luminous ($M_{\rm{UV}}\simeq-22$) galaxies \citep{Stark2017}, suggesting very low metallicity 
gas in some of the most massive galaxies known at $z>7$.  While larger samples are ultimately required to confirm, this is consistent with expectations for evolution toward lower metallicities (at fixed luminosity) between $z\simeq2-3$ and $z\simeq7$.

(5) We compute the fraction of $z\simeq2-3$ galaxies with intense C~{\small III}] emission (EW $>15$~\AA), considering sources that are bright ($M_{\rm{UV}}<-19$) and have large [O~{\small III}]+H$\beta$ EWs ($300-3000$~\AA; median $=760$~\AA) as have been targeted spectroscopically in the reionization era. Even when we consider 
the most extreme [O~{\small III}] emitting galaxies, we find that a very small 
fraction of galaxies exhibit intense C~{\small III}] emission 
($0.013^{+0.031}_{-0.011}$), much lower than the fraction that 
is inferred at $z>5.4$ ($0.20$; \citealt{Mainali2018,Hutchison2019}). This suggests that the intense C~{\small III}] emission seen at $z>6$ is not entirely driven by a shift toward younger stellar populations (i.e., larger [O~{\small III}]+H$\beta$ EW) and 
that evolution toward lower metallicities is also likely 
contributing to the detection rate.  

(6) We quantify the fraction of extreme [O~{\small III}] 
emitters with nebular C~{\small IV} emission in our sample, 
providing a window on the presence of very low metallicity 
stellar populations ($<0.1\ Z_\odot$) at $z\simeq2-3$.  Considering those systems in our sample with [O~{\small III}]+H$\beta$ EW $=1000-3000$~\AA\ and $M_{\rm{UV}}<-19$, we find that only $2$ out of $21$ sources show C~{\small IV} EW $>5$~\AA\, and none of them shows C~{\small IV} EW $>10$~\AA. To account for the possibility that [O~{\small III}] may be weaker at such low metallicities, we also consider another $23$ galaxies with [O~{\small III}]+H$\beta$ EW $=500-1000$.  None shows C~{\small IV} with EW $>5$~\AA. These results stand in contrast to the detection of C~{\small IV} with EW $>20$~\AA\ in two of the first $z>6$ galaxies targeted with $M_{\rm{UV}}<-19$ \citep{Stark2015b,Mainali2017,Schmidt2017}. 
The emergence of such strong C~{\small IV} emission may reflect the 
increased incidence of very low metallicity stars ($<0.1\ Z_\odot$) 
in the reionization-era population.


\section*{Acknowledgements}

The authors thank referee for the helpful suggestions and comments. 
We are grateful for enlightening conversations with Xinnan Du, Alice Shapley, Taylor Hutchison, Michael Maseda, and Kimihiko Nakajima.
DPS acknowledges support from the National Science Foundation through the grant AST-1410155. 
EC acknowledges support from ANID project Basal AFB-170002. 
This work is based on observations taken by the 3D-HST Treasury Program (GO 12177 and 12328) with the NASA/ESA HST, 
which is operated by the Association of Universities for Research in Astronomy, Inc., under NASA contract NAS5-26555.
Observations reported here were obtained from the Magellan Telescopes located at Las Campanas Observatory, Chile, 
and the MMT Observatory, a joint facility of the University of Arizona and the Smithsonian Institution. 
This paper uses data products produced by the OIR Telescope Data Center, supported by the Smithsonian Astrophysical Observatory.
We thank the MMT queue observers, Chun Ly and ShiAnne Kattner, for assisting with our MMT/Binospec observations. 
This work was partially supported  by a NASA Keck PI Data Award, administered by the NASA Exoplanet Science Institute. 
Data presented herein were obtained at the W. M. Keck Observatory from telescope time allocated to 
the National Aeronautics and Space Administration through the agency's scientific partnership with 
the California Institute of Technology and the University of California. 
The Observatory was made possible by the generous financial support of the W. M. Keck Foundation. 
The authors acknowledge the very significant cultural role that the summit of Mauna Kea has always had within 
the indigenous Hawaiian community. We are most fortunate to have the opportunity to conduct observations from this mountain.


\section*{Data Availability}

The data underlying this article will be shared on reasonable request to the corresponding author.



\bibliographystyle{mnras}
\bibliography{UV_spectra}



\appendix


\bsp	
\label{lastpage}
\end{document}